\documentclass{aa}

\usepackage{graphicx}
\usepackage{txfonts}
\usepackage{natbib,twoopt}
\usepackage{upgreek}
\usepackage[colorlinks=true, allcolors=blue]{hyperref}
\usepackage{orcidlink}
\let\orcid\orcidlink

\bibpunct{(}{)}{;}{a}{}{,}
\makeatletter
\renewcommand*\aa@pageof{, page \thepage{} of \pageref*{LastPage}}
\makeatother

\definecolor{grey}{rgb}{0.4,0.5,0.6}
\definecolor{brown}{rgb}{0.65,0.16,0.16}
\definecolor{darkgreen}{rgb}{0.0,0.45,0.0}
\definecolor{darkorange}{rgb}{0.9,0.2,0.0}

\defcitealias{tremaine+77}{TR}
\defcitealias{DiazGimenez+20}{Paper~I}
\defcitealias{DiazGimenez+21}{Paper~II}
\defcitealias{Taverna+22}{Paper~III}
\defcitealias{Taverna+22r}{III}
\defcitealias{zandivarez+23}{Paper~IV}
\defcitealias{taverna+24}{Paper~V}

\begin{document}

   \title{Compact groups from semi-analytical models of galaxy formation}

   \subtitle{VI: Evolution of the two first-ranked galaxies}

   \author{A. Zandivarez\orcid{0000-0003-1022-1547}\inst{1,2}\fnmsep\thanks{\email{ariel.zandivarez@unc.edu.ar}}
          \and
          E. Díaz-Giménez\orcid{0000-0001-5374-4810}\inst{1,2}
          }

   \institute{Universidad Nacional de C\'ordoba (UNC). Observatorio Astron\'omico de C\'ordoba (OAC). C\'ordoba, Argentina
         \and
             CONICET. Instituto de Astronom\'ia Te\'orica y Experimental (IATE). Laprida 854, X5000BGR, C\'ordoba, Argentina
             }

   \date{Received XXX XX 2024 / Accepted XXX XX 2024}

% \abstract{}{}{}{}{} 
% 5 {} token are mandatory
 
  \abstract
   {Compact groups (CGs) of galaxies have proven to be unique environments for studying galaxy interactions. However, there is a need for studies based on statistical evidence 
   to predict how the two main galaxies in these systems went on to develop the properties we observe today.
   }
   {We propose a detailed analysis of the galaxy evolution to disentangle the relationship between the two first-ranked galaxies in CGs throughout their history as a function of the assembly channels of their hosts.
   }
   {Our study was performed from a semi-analytical point of view, using more than $20\,000$ CGs extracted from mock catalogues built from four different semi-analytical models of galaxy formation. We based our analysis on studying the first- (1R) and second-ranked (2R) galaxies in CGs, where the ranking is determined using the galaxy stellar mass as the parameter. 
   }
   {The 1R galaxies have significantly reduced their star-forming capacity over time, reaching a quenching stage and often becoming bulge-dominated or elliptical. Notably, this transformation occurred earlier for 1R galaxies in early-formed CGs (around 5 to 8 Gyrs ago),  while those in recently-formed CGs experienced this change more recently (around 2 to 3 Gyrs ago).
   The analysis of the time evolution of a variant of the Tremaine \& Richstone statistics showed that the 1R galaxy in early-formed CGs began to stand out for its dominant properties around 6 Gyrs ago, almost 5 Gyrs earlier than the 1R inhabiting recently formed CGs. 
   Our merger trees analysis has demonstrated that 1R and 2R galaxies are easily differentiated by their galaxy interactions during their life span. A large majority of the 1R galaxies have experienced at least one major merger event during their life, while we observed this only for a third of 2R galaxies. The 1R galaxies can also display several of these events and most of their last major merger (LMM) events can be described as the addition of a progenitor that is the second most massive galaxy in their surroundings at the time of the merger.  
   }
   {We find that the semi-analytical framework explored in this work describes a scenario where galaxy mergers are the main driving force in shaping the properties of the 1R galaxies in CGs. We note that this scenario is especially intensive when those galaxies inhabit  CGs that had formed early on.}

                
   \titlerunning{Evolution of the two first-ranked galaxies in CGs}
   \authorrunning{A. Zandivarez \& E. D\'iaz-Gim\'enez}
   \maketitle
%
%-------------------------------------------------------------------

\section{Introduction}
The literature on extragalactic astronomy has shown that galaxy evolution is enhanced in dense galaxy systems. Over the years, the search for metrics that help to identify galaxies with extreme evolutionary processes in galaxy systems has been the basis of numerous studies. In particular, one widely used metric is the difference in luminosity between the two brightest galaxies in a system. The use of the luminosity gap between the first- and second-ranked galaxies as an indicator began in the 1970s (e.g. \citealt{sandage+73,geller+76,ostriker+77}) and this approach has been extended over time to study diverse galaxy systems (e.g. \citealt{ponman+94,jones+03,donghia+05,Dariush+07,smith+10,DiazGimenez+11,gozaliasl+14}). This metric has been suggested as a parameter  that is sensitive to the age or assembly history of the galaxies. When the luminosity gap between the two brightest galaxies in the system is considerable, the growth of the dominant galaxy may be due to a dynamically busy past. It is known that mergers between galaxies play an important role in this growth, which can lead to the creation of dominant elliptical galaxies in relatively short periods \citep{barnes+89}, most likely at the expense of their less dominant companions.

The magnitude gap between the two brightest galaxies in a system is not always an accurate indicator of their differences. For instance, there could be a significant magnitude gap between the galaxies, such as a difference of one magnitude or more. However, supposing the magnitude distribution of the brightest galaxies for a particular set of systems is quite broad and includes many potential values of the second brightest galaxy would imply that the two galaxies may not be as different as the magnitude gap suggests. Therefore, reliable statistics should consider not only the magnitude gap, but also the width of the magnitude distributions of these galaxies. Fortunately, extragalactic astronomy has had access to these statistics for quite some time. 
The work of \citet[hereafter \citetalias{tremaine+77}]{tremaine+77} introduced two statistics that weigh the width of the distribution of the brightest and the mean magnitude gap value ($T_1$), as well as the width of the magnitude gap distribution with its mean ($T_2$). These statistics were constructed by specifying that if they exceeded unity, the brightest galaxy has a brightness measurement that is regarded as a random sampling from the luminosity function. When the values of $T_1$ or $T_2$ are below unity and start to be close to 0.7 \citep{mamon87TR}, the first-ranked galaxy is considered to be anomalously bright compared to the second-ranked galaxy.

The first determination by \citetalias{tremaine+77} gave $T_1=0.72$ and $T_2=0.98$ for a sample of clusters with more than 75 members. \cite{loh+06} measured \citetalias{tremaine+77} stats for large red galaxies in the Sloan Digital Sky Survey (SDSS) between 0.12 < z < 0.38 in redshift ranges finding a practically constant behaviour for $T_1$ and $T_2$ with values close to 0.65 and 0.79, respectively. 
Additionally, \cite{lin+10} for a sample of luminous clusters found values of 0.70 and 0.96. \cite{shen+14} obtained a value of $T_1=0.7$ for groups in the SDSS Data Release 7, but could not conclusively determine whether this is due to the first-ranked being brighter than expected or to the satellites being weaker than expected. 
\cite{trevisan+17} found very low values for the \citetalias{tremaine+77} statistics ($T_1$ values less than 0.5, and $T_2$ values less than 0.8) for galaxy groups in the SDSS \citep{yang+07}. 
These findings remain valid as a function of the richness and also as a function of the group mass. 
The authors also observed that the observational $T_1$ and $T_2$ values could not be reproduced with a single Schechter luminosity function for the entire magnitude range, nor a double Schechter luminosity function: one for the centrals and one for the satellites. 
The low $T_1$ and $T_2$ values could only be reproduced with a double Schechter luminosity function and the presence of small-gap groups having two central galaxies, which points to the importance of group mergers in the formation and evolution of these systems. 
On the other hand, \cite{ostriker+19} estimated the $T_1$ values for systems identified in dark matter-only simulations in a Lambda cold dark 
 matter cosmology. They found $T_1$ values even lower than those obtained from observations, concluding that the anomalous gaps are not necessarily related to the luminosity function, but that they are likely due to gravitationally induced mergers caused by dynamical friction.

\cite{ostriker+77} quantitatively demonstrated through numerical predictions that a merger process between the two brightest galaxies can cause the first-ranked galaxy to become brighter, causing a wider gap with the new second-ranked galaxy. Numerical N-body simulations allowed \cite{Mamon87} to conclude that mergers between galaxies can lead to lower values of $T_1$ and $T_2$, generating a more dominant first-ranked galaxy at the expense of the second. After these pioneering works, many other studies support the idea that the growth of the first-ranked galaxy due to mergers is the most probable path (e.g. \citealt{ponman+94,aragon+98,laporte+13,lauer+14,golden-marx+18}).

Therefore, to study the growth of the first-ranked galaxy, we should seek environments that favour interactions between galaxies since they are the most suitable scenarios. It is known that low-mass galaxy groups should be optimal sites for galaxy mergers \citep{Mamon92}. In particular, compact groups (hereafter CGs) of galaxies are relatively isolated systems, where a few galaxies are very close to each other,  making them an ideal environment to study interactions between galaxies. As an example, the iconic Stephan's Quintet CG discovered by \cite{Stephan1877} almost 150 years ago is a clear example of a dynamic scenario for galaxy evolution \citep{moles+97}, which triggered several studies about how galaxy interactions shaped their galaxies and the intra-group medium (e.g, \citealt{emonts+25,arnaudova+24}). The most famous catalogue of CGs is the one developed by \cite{Hickson82}, who visually selected CGs in the plane of the sky by searching for galaxy associations with four or more bright galaxies with high mean surface brightness and without nearby galaxy companions. A decade later, redshift information was added to reduce the contamination due to interlopers \citep{Hickson92}. The overall procedure is popularly known as Hickson's criteria.

Hickson-like CGs have also been a laboratory for studying \citetalias{tremaine+77} statistics. \cite{DiazGimenez+12} obtained values of $T_1 = 0.51$ and $T_2 = 0.70$ for the CG catalogue identified in the Two-Micron All Sky Survey (2MASS). Determinations for CGs identified in synthetic catalogues made from semi-analytical models of galaxy formation yielded similar results although somewhat lower. These observational results were very relevant since they showed lower values for $T_1$ and $T_2$ than for other systems in observational catalogues. 
The authors emphasised that the fact that the catalogue is built on the $K$ band allowed a selection of galaxies practically based on stellar mass. If the segregation by luminosity is due to mergers caused by dynamic friction, stellar masses are a more sensitive variable to detect this effect. That is, selection in the K band should result in lower determinations for the $T$ statistics. In a more global study of the influence of the photometric band on the identification of CGs, \cite{Taverna+16} found that the values of the \citetalias{tremaine+77} statistics are less than unity for all CGs identified in three photometric bands ($K$, $r-SDSS$ and $u-SDSS$ bands). 
In particular, they found that the values of $T_1$ and $T_2$ are higher in the $u$ band and very similar in the $K$ and $r$ bands. However, when systems that only exist in the $K$ band are considered, these present the lowest possible values for $T_1$ and $T_2$, while the highest values were obtained for those systems that only existed in the $u$ band. These results suggest that the photometric band used to identify CGs can influence the values of the \citetalias{tremaine+77} statistics and, therefore, the interpretation of the merger history within the groups. 

To deepen our understanding of how a dominant galaxy in CG can form and exhibit differences from its immediate companion, it is essential to investigate the evolutionary history of these pairs over time. A key approach for conducting a statistically reliable analysis of this phenomenon is to perform numerical studies based on synthetic catalogues. This tool allows for exploring the complex interactions and developments within these galaxy pairs effectively. Over the last five years, we conducted a series of studies using several mock galaxy catalogues created from numerical simulations and semi-analytical models (SAMs) of galaxy formation to analyse different aspects of Hickson-like galaxy groups (CGs). This series of studies aimed to deepen our understanding of the nature and behaviour of CGs in the cosmos.
Our investigations included analysing the frequency and nature of  CGs (\citealt{DiazGimenez+20}, \citetalias{DiazGimenez+20}), exploring their various assembly channels (\citealt{DiazGimenez+21}, \citetalias{DiazGimenez+21}), evaluating the performance of the Hickson-like automatic algorithm utilised in flux-limited catalogues (\citealt{Taverna+22}, \citetalias{Taverna+22}), assessing the time evolution of the properties of galaxies in CGs based on their assembly channels (\citealt{zandivarez+23}, \citetalias{zandivarez+23}), and predicting the location of CGs characterised by specific assembly channels within the large-scale structure of the Universe (\citealt{taverna+24}, \citetalias{taverna+24}).

In this work, we continue with the series using the semi-analytical CGs. This time, we concentrate on the two first-ranked galaxies in CGs when the ranking is performed using the galaxy stellar mass. Our objective is to understand how the first-ranked galaxy evolves into its current state, how this process compares to the evolution of its second-ranked companion, and whether these evolutionary paths depend on the assembly channel of the CG that hosts them.

The layout of this work is as follows. In Sect.~\ref{sec:samples}, we present the mock catalogues used in this work and the sample of CGs. We also detail the procedure to classify CGs according to their assembly channel. In Sect.~\ref{sec:results}, we describe all the analyses performed on the two first-ranked galaxies in CGs. We analysed the morphological transformations and star formation capabilities of the two top-ranked galaxies over time. Additionally, we calculated the evolution of the \citetalias{tremaine+77} statistics and conducted a detailed examination of the major merger events that shaped the evolutionary paths of these galaxies. We summarise and discuss our results in Sect.~\ref{sec:conclusions}.  

%--------------------------------------------------------------------
\section{Samples}
\label{sec:samples}
Throughout this work, we adopted the samples previously analysed in \cite{taverna+24}. In the following subsections, we briefly describe the methods to build these samples.

\subsection{Mock catalogues}
The mock catalogues are lightcones built from the Millennium I Simulation \citep{Springel+05, Guo+13} with different cosmologies. These simulations of dark matter particles are combined with different SAMs\footnote{\url{http://gavo.mpa-garching.mpg.de/Millennium/}} to assign synthetic galaxies following different recipes. The SAMs used in this work are G11 \citep{Guo+11}, G13 \citep{Guo+13}, H20 \citep{Henriques+20}, and A21 \citep{Ayromlou21}. In Table~\ref{tab:sams}, we quote the cosmological parameters of the simulations used for each SAM.

These publicly available catalogues are used to build lightcones following the procedure outlined in \cite{jpas}. Using snapshots of different cosmic epochs from the galaxy catalogues, we mimic the evolution of structures in the universe by stacking snapshot slices of a given width. The galaxy rest-frame absolute magnitudes obtained in the SAMs plus a K-decorrection procedure \citep{DiazGimenez+18} are used to estimate the observer-frame apparent magnitudes in the lightcone.
The resulting mock lightcones are all-sky surveys with redshifts less than 0.2 and a limiting observer frame apparent magnitude of 17.77 in the SDSS r-band. Following previous works, we only included galaxies with stellar masses larger than $\sim 10^9 \mathcal{M}_\odot$.  

\subsection{CG identification}
The CG samples are identified using the Hickson-like finding algorithm devised by \cite{DiazGimenez+18}.
The algorithm searches for galaxy systems that fit the set of constraints on membership, compactness, relative isolation, velocity concordance, and flux limit of the brightest group galaxy (BGG).
Specifically, the overall criteria require: 
\begin{itemize}
    \item Population: There are three or more galaxy members within a three magnitude range from the brightest.
    \item Flux limit of the BGG: the BGG of the system has to be at least three magnitudes brighter than the catalogue magnitude limit. This constraint is included to maximise completeness \citep{Prandoni+94,DiazGimenez&Mamon10}.
    \item Velocity concordance: galaxy members are within $1\,000 \rm \, km\,s^{-1}$ from the median velocity of the system \citep{Hickson92}. 
    \item Compactness: the surface brightness of the system in the $r$-band is less than $26.33 \, \rm mag \, arcsec^{-2}$ (this is the 26 limit defined by \citealt{Hickson82} in the $R$ band but modified for the $r$ SDSS band). 
    \item Relative isolation: there are no other bright galaxies (in a three-magnitude range from the brightest, nor brighter) within three times the radius of the minimum circle that encloses the galaxy members.
\end{itemize}

Since semi-analytical galaxies are point objects, to mimic observational constraints, we included a recipe to deal with the possible blending of galaxies. Following \cite{Lange+15}, we used the galaxy stellar mass estimated by the SAM to assign half-light radius to each dimensionless mock galaxy. 
Then, we considered two mock galaxies as blended if their radii overlap in projection. 
%After blending, we refer to the number of members as \textit{observable} number of members \euge{volver aqui: alguna vez se usa esta definicion de observables?}. 
In Table~\ref{tab:sams}, we quote the number of CGs identified in each lightcone. 

Since Hickson-like algorithms identify a fraction of not physically dense systems due to the chance alignments of galaxies along the line of sight \citep{McConnachie+08,DiazGimenez&Mamon10}, we excluded these fake systems from our samples: we discard those systems with a maximum 3D inter-galaxy separation larger than 1 Mpc $h^{-1}$ at present, as well as those without galaxy pairs separated by less than 200 kpc $h^{-1}$ in 3D real space (see column 7 of Table~\ref{tab:sams}).

%-----------------------------------------------------
\begin{figure*}
   \centering
   \includegraphics[width=0.49\textwidth]{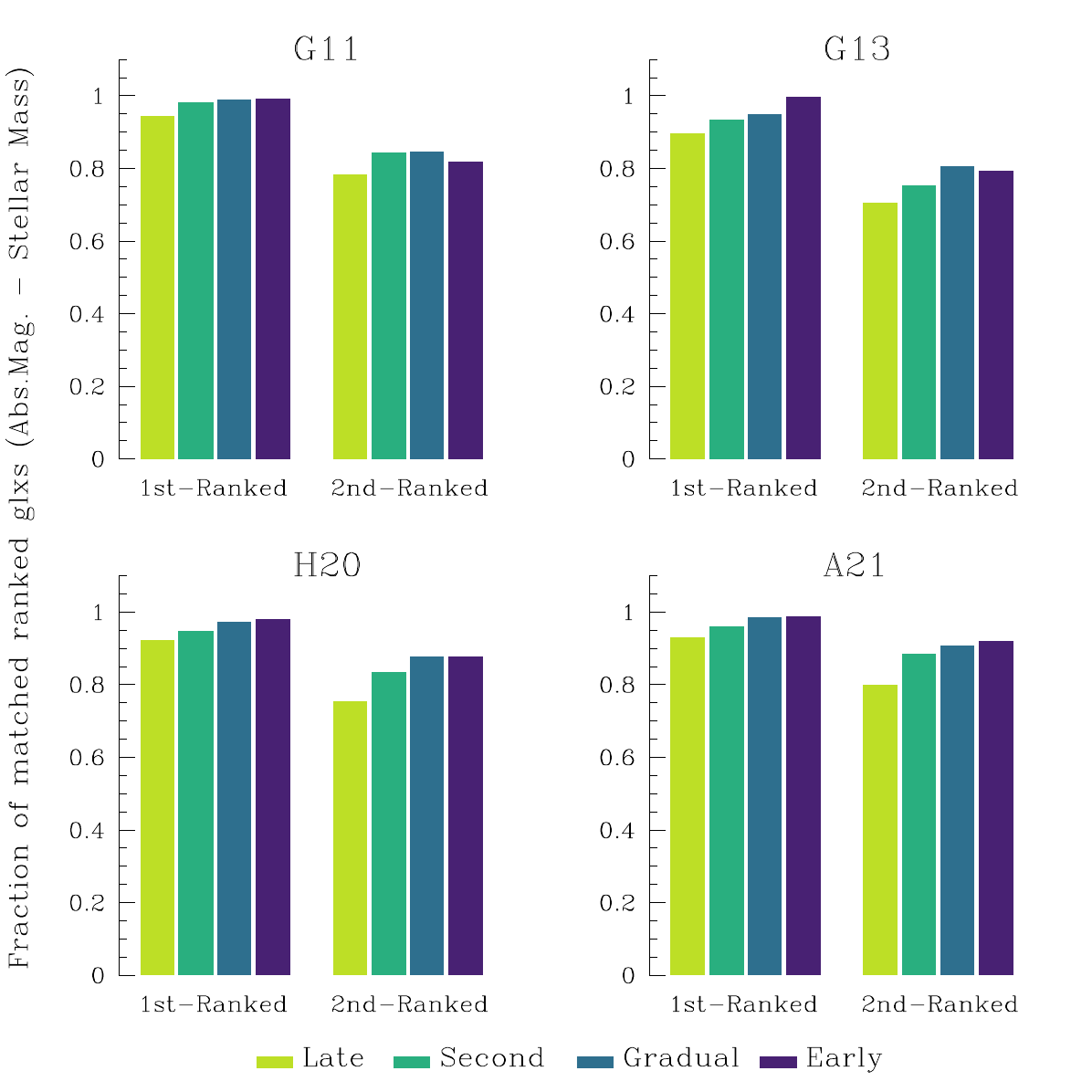}
%/big4/users/euge/SAM_SED/env_channels/env_channels_prog/props.sm percent   
   \includegraphics[width=0.49\textwidth]{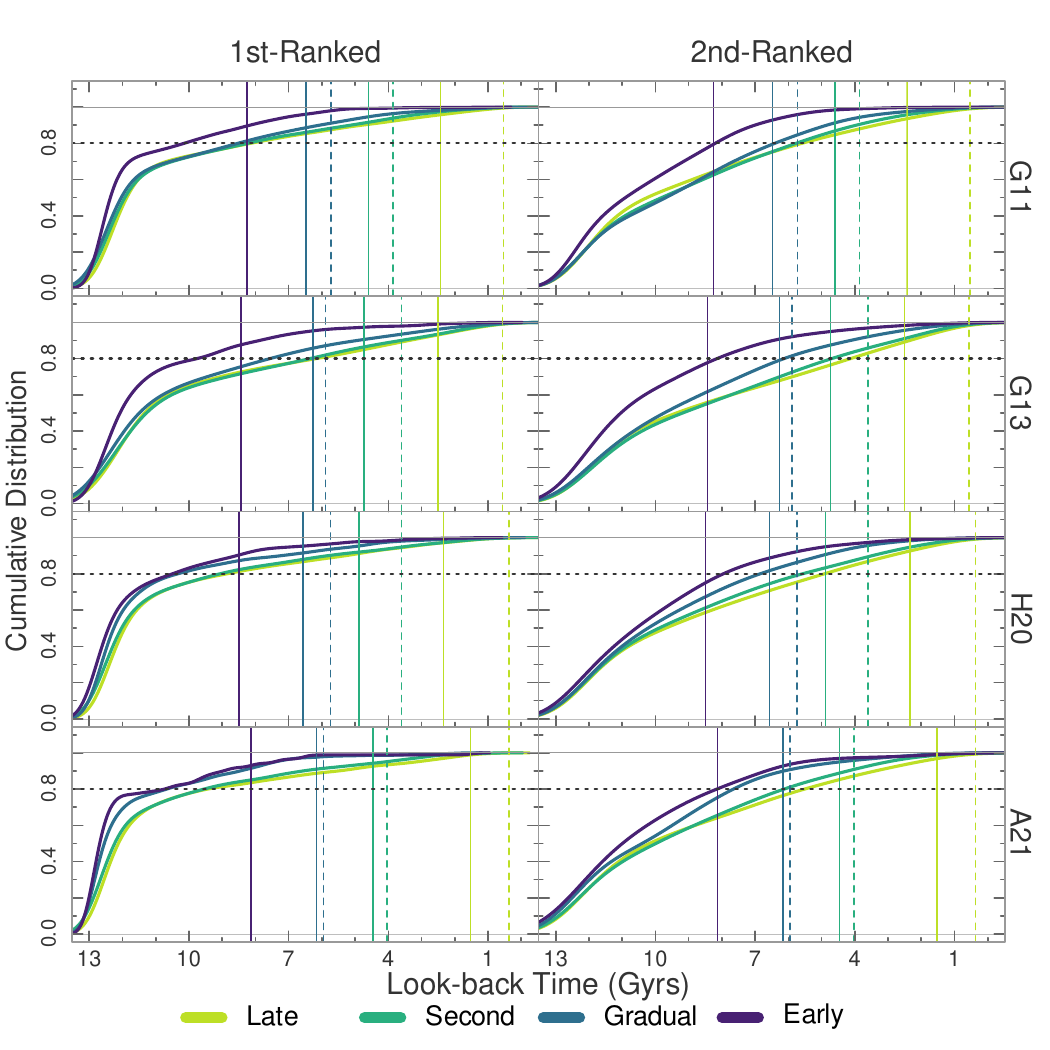}
%/big4/users/euge/SAM_SED/env_channels/env_channels_prog/plotTimes.R
      \caption{Preliminary comparison between the two first-ranked galaxies in CGs selected at $z=0$. Left plots: Fraction of the two first-ranked galaxies in CGs whose rankings in absolute magnitude and stellar mass match. The bar plots are shown as a function of the SAMs and the assembly channels of their CG hosts. Right plots: Cumulative distributions of look-back times since the two first-ranked galaxies selected by stellar mass at $z=0$ achieved their current ranking (only CG members at $z=0$ are used). Each distribution is shown as a function of the assembly channel of the host CG. Vertical solid lines represent the median time for the two first-ranked galaxies to be considered under mutual gravitational interaction. Dashed lines indicate the median first pericenter time of the last-arriving CG member (key galaxy).}
         \label{fig:1}
\end{figure*}
%-----------------------------------------------------

\begin{table}
\setlength{\tabcolsep}{1pt}
\begin{center} 
\caption{Compact groups from mock lightcones \label{tab:sams}}
\begin{tabular}{clccccccccccc}
%\begin{tabular}{clccccc}
\hline
\hline
\multicolumn{1}{c}{\small SAM} &\multicolumn{3}{c}{\small Simulation} & &  \multicolumn{7}{c}{\small Mock} \\ 
\cline{2-4} \cline{6-12}
{\small acronym} & \multicolumn{1}{c}{\small name} & {\small $\Omega_{\rm m}$} & {\small $h$} && {\small CG} & {\small CG$^{3+4}$} & {\small CG$^{3+4}_F$} & {\small L} & {\small S} & {\small G} & {\small E} \\
\multicolumn{1}{c}{\small (1)\ \ } & \multicolumn{1}{c}{\small (2)} &\multicolumn{1}{c}{\small (3)} & {\small (4)} && {\small (5)} & {\small (6)} & {\small (7)} &\multicolumn{4}{c}{\small (8)}\\
\hline
\small G11 & \small \ WMAP1  & \small 0.25 & \small 0.73 & & \small 7850 & \small 7129 & 
 \small 912 & \small 2180 & \small 1671 & \small 1621 & \small 745 \\
\small G13 & \small \ WMAP7  & \small 0.27 & \small 0.70 & & \small 5233 & \small 4901 & 
 \small 701 & \small 1816 & \small 1197 & \small 870 & \small 317 \\
\small H20 & \small \ Planck & \small 0.31 & \small 0.67 & & \small 5881 & \small 5487 & 
 \small 976 & \small 2011 & \small 1344 & \small 827 & \small 329 \\
\small A21 & \small \ Planck & \small 0.31 & \small 0.67 & & \small 2927 & \small 2794 & 
 \small 712 & \small 1074 & \small 625 & \small 296 & \small 87 \\
\hline
\end{tabular}  
\end{center} 
\parbox{\hsize}{\noindent \small Notes:
(1): SAM: G11
\citep{Guo+11}, G13 \citep{Guo+13}, H20 \citep{Henriques+20} and A21 \citep{Ayromlou21}; 
(2): cosmology: WMAP1 \citep{Spergel+03}, WMAP7 \citep{Komatsu+11}, Planck \citep{Planck+16};
(3): dimensionless matter density parameter;
(4): dimensionless $z$=0 Hubble constant;
(5): number of compact groups in the lightcones;
(6): number of compact groups with exactly three or four galaxy members;
(7): a subsample of (6) classified as 'fake' systems;
(8): subsamples of (6) split according to their assembly channel (late, second, gradual, and early). \\
}
\end{table}

\subsection{CG assembly channels}
\citet[hereafter \citetalias{DiazGimenez+21}]{DiazGimenez+21} outlined a procedure to classify CGs into different assembly channels using the information provided by the merger trees of the CG galaxy members.

Following the main branch of each galaxy member, $j$, we computed the physical 3D distance to the stellar mass centre of the group in each output, $i$, of the simulation, $p_j(t_i)$, where $t_i$ is the look-back time of the snapshot $i$. 
Then we extracted from each profile:
\begin{itemize}
    \item $t_{\rm 1p}$: the look-back time of the galaxy first pericenter passage;
    \item $n_{\rm p}$: the number of pericentres passages (each one clearly defined between the four nearest snapshots);
    \item $r_{01}$: the ratio between the $p_j(t_i)$ values in $t_0$ (latest snapshot) and the immediately previous output, $t_1$.
\end{itemize}

From the profiles determined for each CG member, \citetalias{DiazGimenez+21} choose one in particular, corresponding with the so-called 'key' galaxy in the system. The 'key' galaxy is the CG member with the latest arrival to the system. 
This characteristic is given to the galaxy with the lowest $n_{\rm p}$. In case there is more than one galaxy, the 'key' label is given to the one with the latest $t_{\rm 1p}$ (provided that their $t_{\rm 1p}$ values are separated by more than 1 Gyr). If the similar galaxy properties situation persists, then the 'key' galaxy is chosen based on the highest value of $p_j(t_{\rm 1p})$ at its first pericenter.
After selecting the 'key' galaxy, \citetalias{DiazGimenez+21} established a procedure originally devised to define four different CG assembly channels for CGs with four galaxy members. This procedure was recently used by \citet[hereafter \citetalias{taverna+24}]{taverna+24} to classify CG with three and four galaxy members. The procedure performed in both works is as follows:
\begin{itemize}
    \item If the key galaxy has $t_{\rm 1p} \le 7.5$ Gyr, then there are three possibilities:
    \begin{itemize}
        \itemindent=0pt
        \item  {\tt Late assembly}: the key galaxy has just arrived in the CG (hereafter {\tt Late}). The key galaxy exhibits $n_{\rm p}=0$ OR [$n_{\rm p}=1$ AND $r_{01}>1$].
        \item {\tt Late second pericentre}: the key galaxy is on its second passage in the CG (hereafter, {\tt second}), namely, when [$n_{\rm p}=1$ AND $r_{01}\leq 1$] OR [$n_{\rm p}=2$ AND $r_{01} > 1$].
        \item {\tt Gradual contraction}: the key galaxy has already completed two or more orbits, becoming gradually closer with each orbit. This behaviour (hereafter, {\tt gradual}) is obtained when [$n_{\rm p}=2$ AND $r_{01} \leq 1$] OR $n_{\rm p}>2$.
    \end{itemize}
    \item If the key galaxy has an early arrival at the CG, namely, $ t_{\rm 1p} \ge 7.5$ Gyr:
    \begin{itemize}
        \itemindent=0pt
        \item {\tt Early assembly}: all the galaxies have been together from an early epoch (hereafter, {\tt early}).
    \end{itemize}
\end{itemize}

The application of this procedure to CGs with three and four galaxy members allows us to include more than $90\%$ of the mock CG samples. The number of CGs with only three and four members (CG$^{3+4}$) is quoted in Table~\ref{tab:sams} as well as the number of those CGs classified in each assembly channel.

%-------------------------------------------------------
\begin{figure*}
   \centering
   \includegraphics[width=0.48\textwidth]{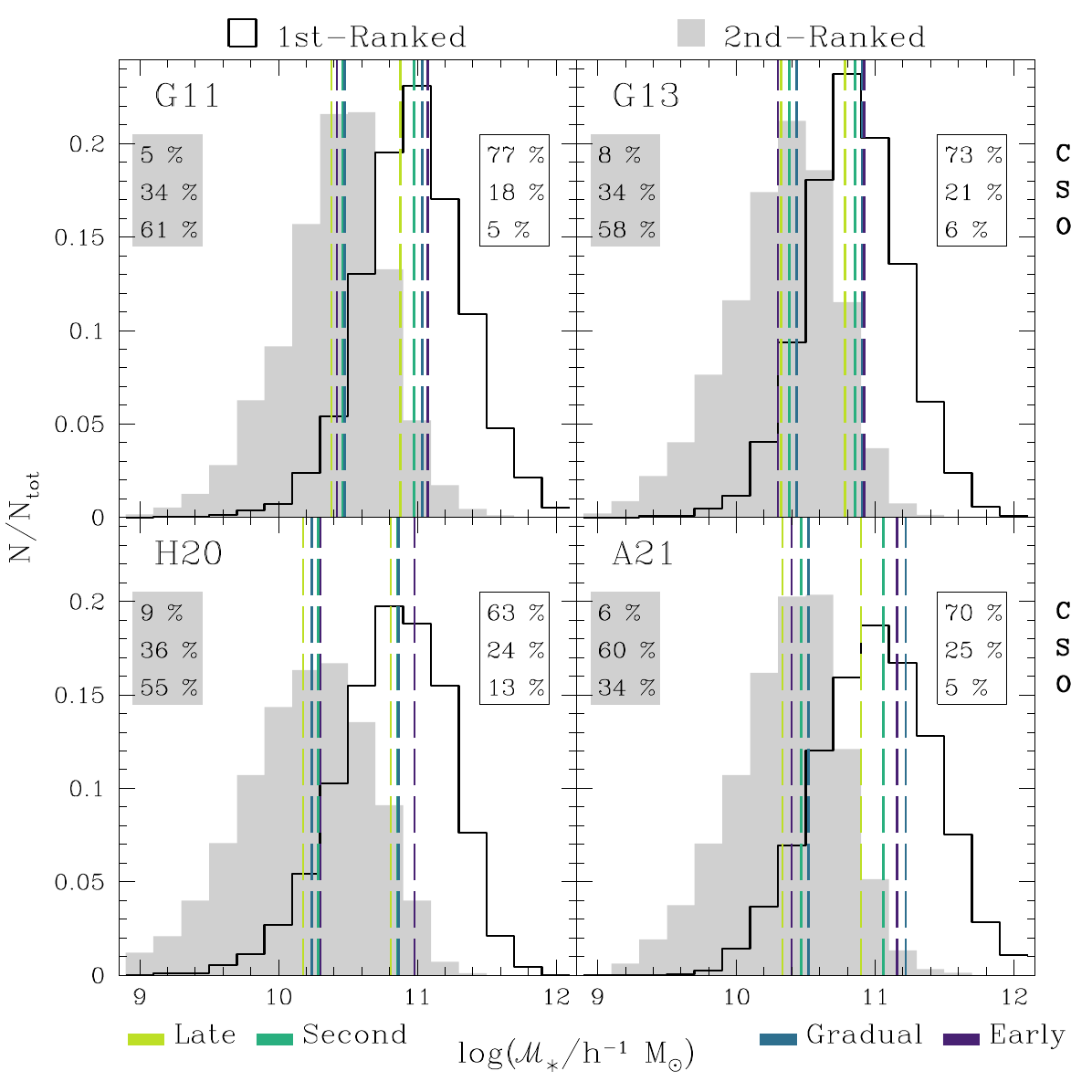}
%/big4/users/euge/SAM_SED/env_channels/env_channels_prog/props.sm masas
\includegraphics[width=0.49\textwidth]{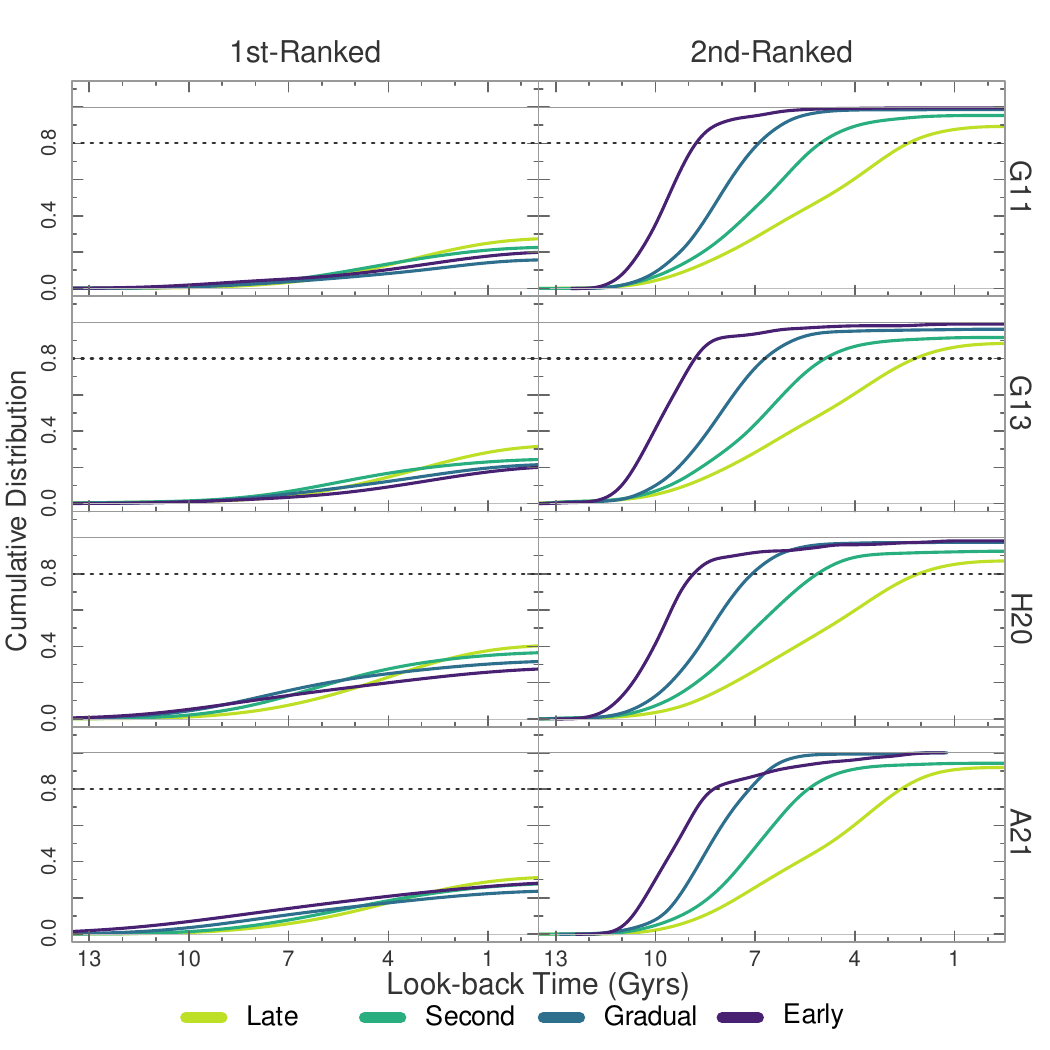}
%/big4/users/euge/SAM_SED/env_channels/env_channels_prog/plotTimes_tipo.R
      \caption{Basic properties of the two first-ranked galaxies by stellar mass in CGs. Left plot: Stellar mass distribution for the two first-ranked galaxies in CGs at $z=0$ for each SAM. Vertical dashed lines indicate the median stellar masses for the samples split according to the CG assembly channel. Inset numeric labels indicate the percentage of different types of galaxies (C: central, S: satellite, O: orphan) for each ranked galaxy sample. Right plot: Cumulative distribution of the look-back times at which the different ranked galaxies have undergone a galaxy-type transition from central to satellite/orphan. Solid and dashed horizontal lines at 1 and 0.8, respectively, are included as references.
              }
         \label{fig:2}
\end{figure*}
%------------------------------
\section{The two first-ranked galaxies in CGs}
\label{sec:results}
In this section, we describe how we performed the main analysis of the two first-ranked galaxies inhabiting CGs. In Sect.~\ref{sec:preliminary}, we introduce the samples of ranked galaxies and their main characteristics, while in Sect.~\ref{sec:evol1}, we study the morphological transformation suffered by the ranked galaxies and their ability to form stars over time. In Sect.~\ref{sec:tr}, we analyse the evolution of the \citetalias{tremaine+77} statistics for CGs, while the major mergers that occurred in the ranked galaxies are studied in Sect.~\ref{sec:mergers}.

\subsection{A preliminary analysis of the selected ranked galaxies}
\label{sec:preliminary}
The analysis of the two most dominant galaxies in a galaxy system is usually approached in the literature by using two galaxy properties to perform the ranking: the absolute magnitude and the stellar mass. It is known that both properties are related, however, we know from previous works \citep{DiazGimenez+12,Taverna+16} that when the photometric bands are more related to a selection by stellar mass, the \citetalias{tremaine+77} statistics show clearer results. On the other hand, the evolution of the variation of the stellar mass with time has a more direct interpretation based on the different processes that can cause changes. 
Thus, we performed our ranking based on the stellar mass of the galaxies. Nevertheless, it is worth noting that, generally (and at least for the semi-analytical samples used here), the ranking in stellar mass usually matches the ranking in absolute magnitude. 
In the left plots of Fig.~\ref{fig:1}, we display the fraction of the first-ranked and second-ranked (hereafter, 1R and 2R, respectively) in stellar mass that are also 1R and 2R in absolute magnitude at present ($z=0$). 
The fractions indicate that the matching using either absolute magnitude or stellar mass is between 90 to 100\% for the 1R,  
while for the 2R galaxies goes from 71 to 92\%. Hence, the study using stellar mass ranking is closely related to one conducted using a ranking based on absolute magnitudes.  

Since the ranking is performed based on the information at $z=0$, we might wonder how long these galaxies held their ranking. 
In the right plots of Fig.~\ref{fig:1}, we show the cumulative distribution of look-back times since the 1R and 2R galaxies at $z=0$ achieved their ranking. 80\% of the 1R (2R) galaxies have held that category during roughly the last 8, 6.5, 8, and 10 (6, 4, 5 and 5.5) Gyrs for G11, G13, H20, and A21, respectively. Regardless of the ranking, these times occur from 1 to 4 Gyrs earlier for galaxies inhabiting CGs described by an {\tt early} assembly channel. 

Dashed vertical lines in the right plots of Fig.~\ref{fig:1} indicate the median arrival time of the key galaxy (the last to arrive in the system), while solid vertical lines indicate the median time since the two first-ranked galaxies can be considered relatively close\footnote{This time of interaction is estimated from the $t_{1p}$ and $t_{200}$ (the first time the galaxy is closer than 200 kpc $h^{-1}$ from the centre of mass of the CG) of both ranked galaxies. The time of interaction for each ranked galaxy is $t_{1p}$ unless $t_{200}$ occurs at least 2 Gyrs earlier than $t_{1p}$, in which case the time of interaction is set to be $t_{200}$. Then, by comparing these times computed for 1R and 2R galaxies, the definitive interaction time is the latest (closest to $z=0$).}. These times allow us to conclude that most of the 1R and 2R galaxies reached their category before the group could be considered fully assembled and before the two first-ranked galaxies began interacting closely.

Figure~\ref{fig:2} show some basic properties of the 1R and 2R galaxies in CGs. In the left plots, we show the stellar mass distributions. The 1R galaxies have median logarithmic stellar masses of  10.97, 10.84, 10.85, and 11, while 2R galaxies display medians of 10.43, 10.36, 10.22 and 10.40, both for G11, G13, H20, and A21, respectively. These values imply that, regardless of the SAM, the 1R galaxies are approximately four times more massive than the 2R galaxies. 
When considering the assembly channels, there is a tendency of 1R galaxies to display a progressive increasing median stellar mass from {\tt late} to {\tt early} CGs, but that behaviour is not observed in the 2R counterparts.

%-----------------------------------------------------
\begin{figure*}
   \centering
   \includegraphics[width=0.49\textwidth]{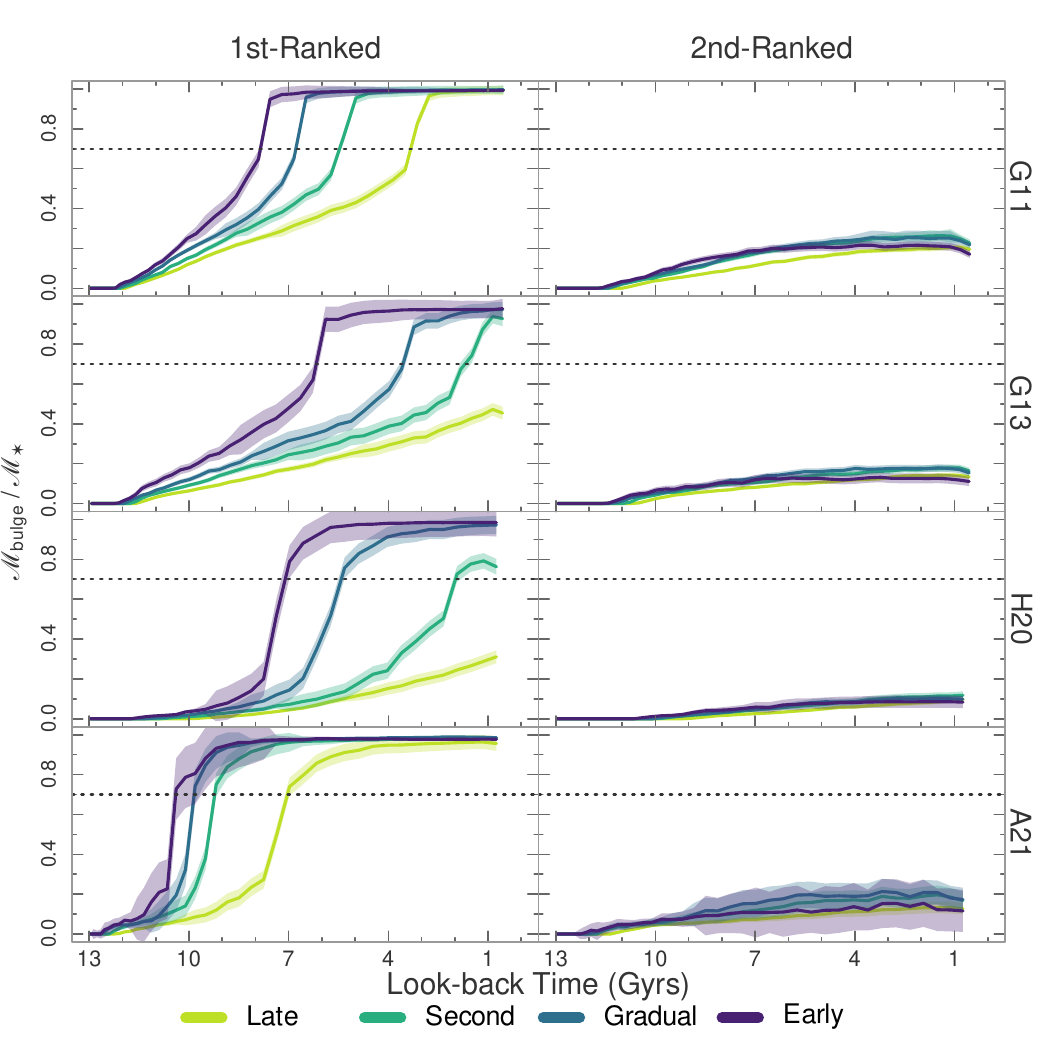}
%/big4/users/euge/SAM_SED/env_channels/env_channels_prog/plotMorf2.R
   \includegraphics[width=0.49\textwidth]{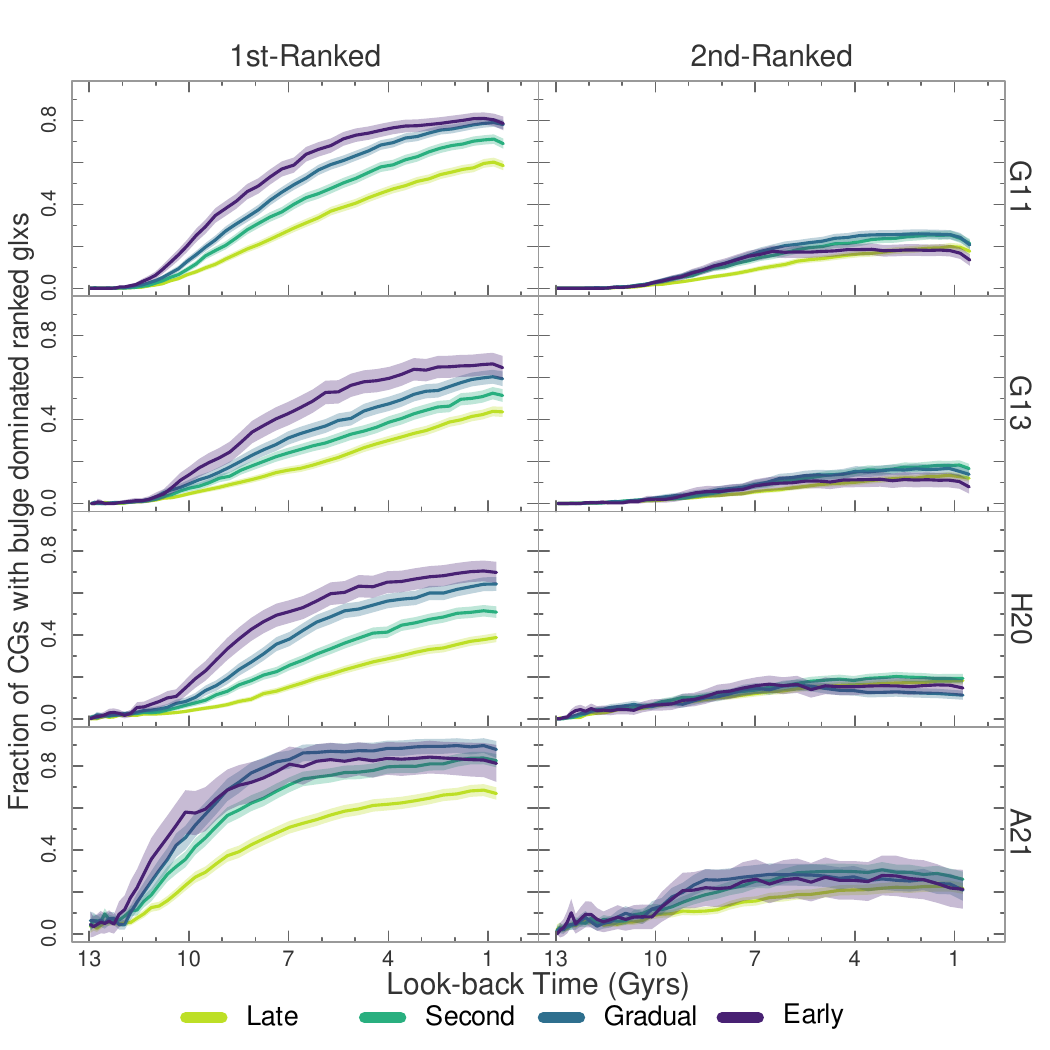}
%/big4/users/euge/SAM_SED/env_channels/env_channels_prog/plotMorf.R
      \caption{Evolution of bulge-dominated phase in the two first-ranked galaxies in CGs as a function of the CG assembly channel. Left plot: Time evolution of the median bulge-to-total stellar mass ratio for the ranked galaxies in CGs. Error bars are the 95\% confidence interval for the medians, $\pm 1.58 \, {\rm IQR}/\sqrt{N}$, with IQR the interquartile range \citep{chambers+83}. These error bars are included only for the seek of comparisons between medians and are not representative of the distribution of values around the medians. The dashed horizontal line at 0.7 indicates the transition to a bulge-dominated (elliptical) morphology. Right plot: Time evolution of the fraction of CGs with bulge-dominated ranked galaxies. Error bars are 95\% binomial confidence interval, $\pm 1.96 \sqrt{f(1-f)/N}$, with $f$ as the fraction.
              }
         \label{fig:3}
\end{figure*}
%-----------------------------------------------------

In the inset labels of the left plots of Fig.~\ref{fig:2}, we included the percentage of the galaxy types in each set of ranked galaxies. In the SAMs used here, galaxies are classified into three types: 'centrals' for galaxies that reside in the central subhalo of a friends-of-friends (FOF) halo; 'satellites', with a not central subhalo within a given FOF; and orphans, which are satellite galaxies without a detectable dark-matter subhalo since it would have been tidally disrupted. In the last case, since tidal stripping is a highly effective process within a halo \citep{delucia+04}, orphan galaxies are a common subproduct in SAMs.   
The 1R galaxy sample is composed on average of 70\%  of centrals, 22\% of satellites, and 7\% of orphans. On the other hand, the 2R galaxy sample is composed of 7\% of centrals, 35\% of satellites and 58\% of orphans for G11, G13, and H20. This latter average result is different for A21 since the population of satellites in CGs is dominant compared with orphan galaxies. These results are similar to those reported in \citetalias{zandivarez+23}, where the entire galaxy population of CGs was analysed based on stellar mass. They found that central galaxies dominate the most massive bins, while satellites and orphans dominate the lower mass ranges (refer to their Fig.~3).

To complete the picture for the two first-ranked galaxies, in the right plots of Fig.~\ref{fig:2}, we show the cumulative distribution of look-back times at which each galaxy has lost their category of central. Regardless of the SAM, the 1R galaxies show a very similar behaviour where $\sim 30\%$ gradually lost their central feature.
On the other hand, the time when most of the 2R galaxies have lost their central feature is a function of the CG assembly channel. While {\tt late} CGs show that $\sim 80\%$ of their 2R galaxies have ceased to be central $\sim 2$ Gyrs ago, the 2R galaxies in {\tt early} CGs are not longer centrals since 7 Gyrs earlier than their {\tt late} counterparts.

\subsection{Evolution of the properties of the two most massive galaxies} 
\label{sec:evol1}
We explore the evolution of two galaxy properties related to their morphology and their efficiency in the formation of stars.
We estimated the bulge-to-total stellar mass ratio, ${\cal M}_{\rm bulge}/{\cal M}_{\star}$, for each ranked galaxy in CGs. Morphological types are usually inferred from this ratio when using SAMs. In particular, some authors have shown that a lower limit of 0.7 in this ratio leads to a reasonable reproduction of the fraction of elliptical galaxies as a function of the stellar mass obtained from observable data (see \citealt{Bertone+07,Guo+11}). 
In this work, we  interchangeably use bulge-dominated or elliptical designations to refer to galaxies with a bulge-to-total mass ratio greater than the previously established limit.

In the left panels of Fig.~\ref{fig:3}, we show the median bulge-to-total mass ratio as a function of look-back time for the 1R and 2R galaxies inhabiting CGs with different assembly channels. Using the reference line at 0.7, we observe that 1R galaxies start their transition to a dominant population of elliptical galaxies in a progressive sequence related to their CG assembly channels. 
The 1R galaxies inhabiting {\tt early} CGs 
have been predominantly\footnote{the median of the ratio of the sample is close to 1} elliptical galaxies during the last 8 Gyrs for G11 and A21, while the 1R galaxies became predominantly elliptical between 5 to 6 Gyrs ago for G13 and H20. 
Meanwhile, the 1R galaxies in {\tt late} CGs display a dominant elliptical population during the last 2 to 3 Gyrs in G11 and A21, 
however, they never end up becoming the dominant population for G13 and H20. 
When analysing the evolution of the 2R galaxies, a very different behaviour is observed. The elliptical galaxies never dominate the 2R sample nor display a distinction between the assembly channels.

In the right plots of Fig.~\ref{fig:3}, we show the evolution of the fraction of CGs with bulge-dominated ranked galaxies. 
These curves quantify the progressive increment of bulge-dominated ranked galaxies in CGs, showing the clear dependence of the 1R as a function of the assembly channel of the CG they inhabit.
For instance, it can be seen that $\sim 60\%$ of {\tt late} CGs in G11 at 1 Gyr have a bulge-dominated 1R galaxy, which implies that the bulge-to-total ratio distribution is very skewed, with a median at 1 (shown in the left plot) and a long tail toward lower values, and with $\sim 40\%$ of the galaxies having ratio values below 0.7.
In general, between 65 to 80\% of the 1R galaxies of {\tt early} CGs at $z=0$ are bulge-dominated, while 40 to 65\% of the 1R in {\tt Late} CGs are bulge-dominated. Although similar behaviours are observed as a function of the CG assembly channels for all SAMs, 
it is worth noting that the rate of change is faster in G11 and A21 than in G13 and H20. 
Therefore, lower rates of change for G13 and H20 produce the lowest fractions of CG with galaxies dominated by bulges in 1R galaxies at $z=0$. 
On the other hand, regardless of the SAM, only 10 to 20\% of the 2R galaxies at $z=0$ are bulge-dominated.

%-----------------------------------------------------
\begin{figure*}
   \centering
   \includegraphics[width=0.49\textwidth]{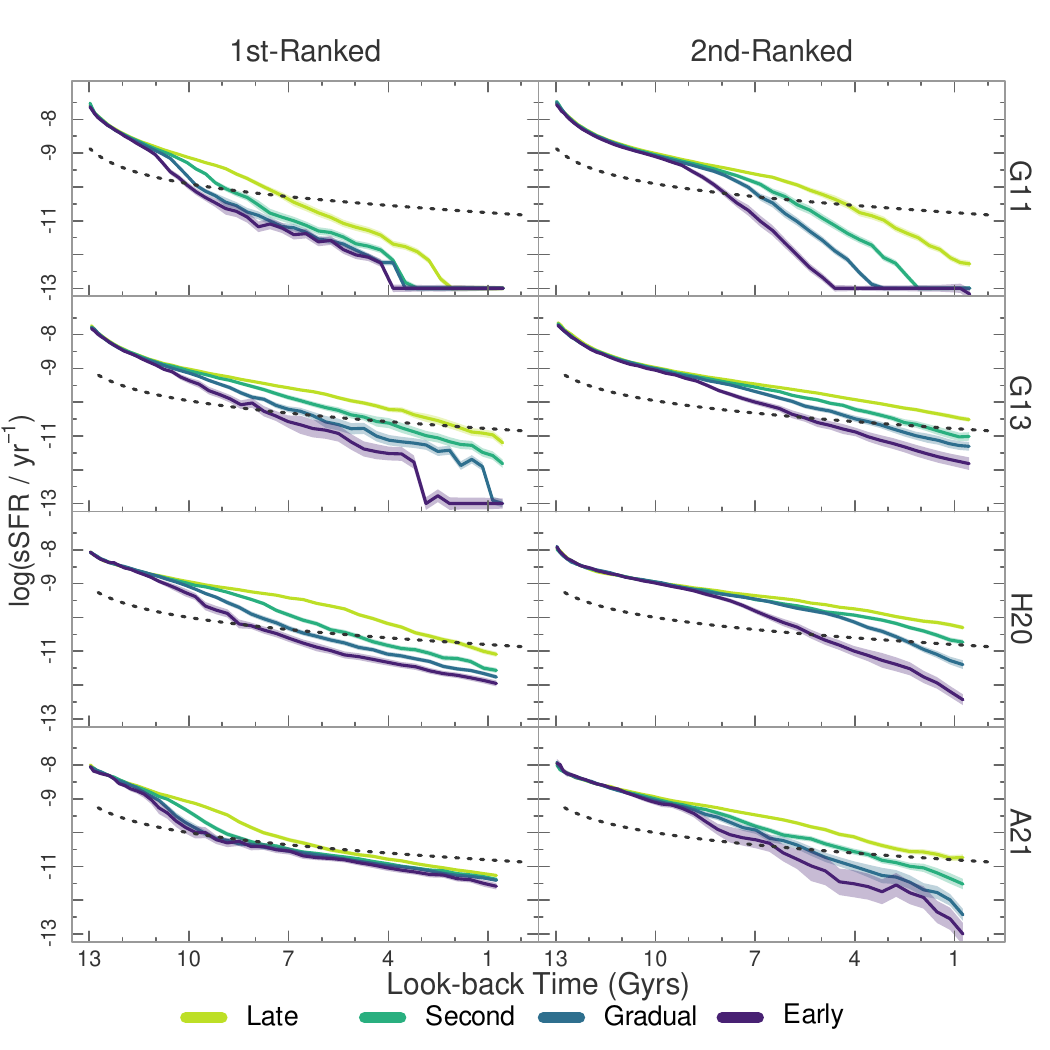}
%/big4/users/euge/SAM_SED/env_channels/env_channels_prog/plotSsfr2.R
   \includegraphics[width=0.49\textwidth]{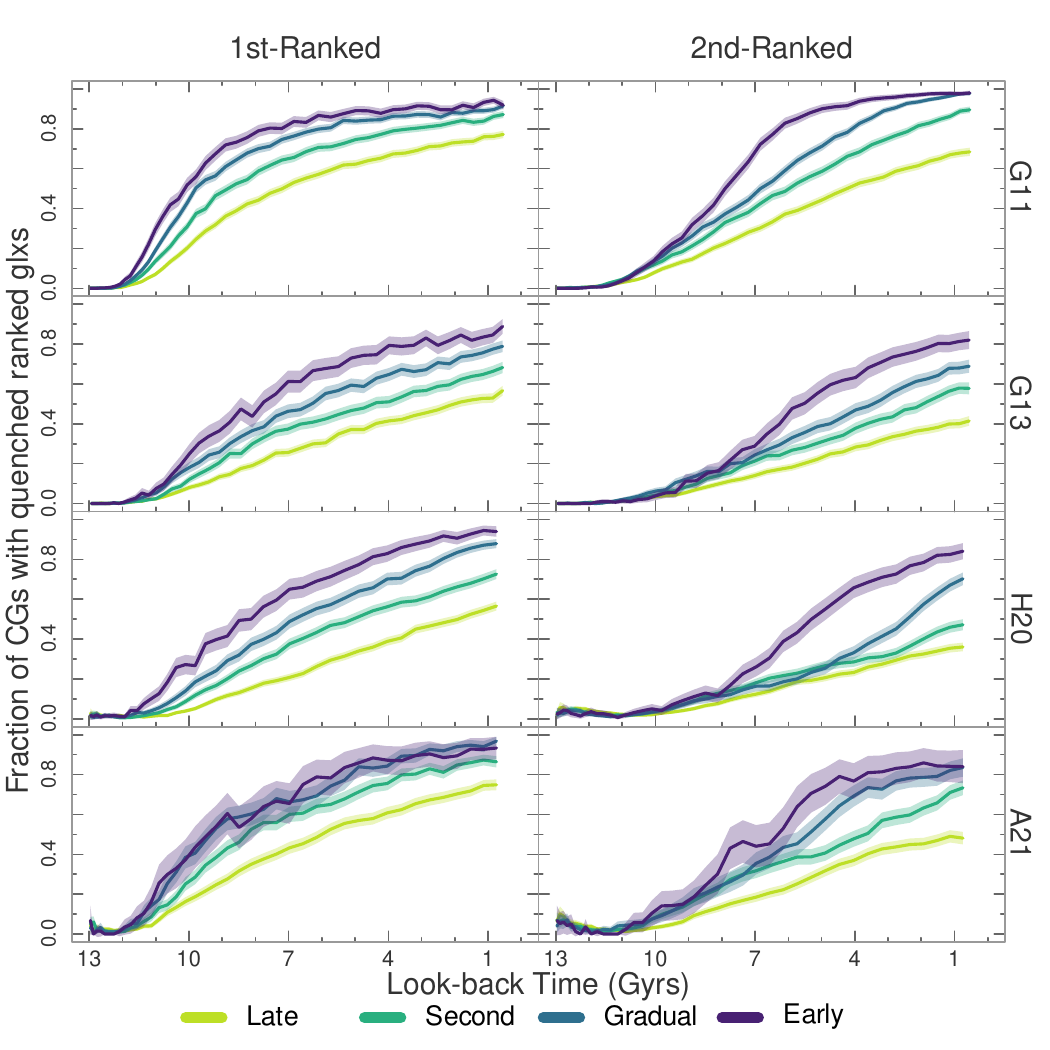}
%/big4/users/euge/SAM_SED/env_channels/env_channels_prog/plotSsfr.R
      \caption{Evolution of the specific star formation rate in the two first-ranked galaxies in CGs as a function of the CG assembly channel. Left plot: Time evolution of the median sSFR for the ranked galaxies in CGs. The black dashed curve indicates the transition to a quenched state (see text). Right plot: The time evolution of the fraction of CGs with quenched ranked galaxies. For both plots, error bars are computed as in Fig.~\ref{fig:3}.
              }
         \label{fig:4}
\end{figure*}
%-----------------------------------------------------

Another way to approach the different galaxy transformations the two first-ranked galaxies have undergone during their history is by analysing their ability to form stars. 
In \citetalias{zandivarez+23}, we  showed that galaxies in CG quartets typically exhibit a lower specific star formation rate (sSFR) than the general population of galaxies, producing a larger fraction of quenched galaxies at $z=0$ in CGs. Here, we focus on the two first-ranked galaxies for a larger sample of CGs (with three and four galaxy members).
In the left plots of Fig.~\ref{fig:4}, we show the evolution of the median sSFR of the 1R and 2R galaxies in CGs. 
Regardless of the SAM and the ranking, the sSFR varies differently according to the assembly channel of the CG they inhabit. Galaxies in {\tt early} CGs suffer a faster decrement of their sSFR than those inhabiting {\tt late} CGs. In addition, the rate of this decrement is a function of the galaxy ranking. The 1R galaxies display a more rapid rate of decrement than the 2R galaxies. 

A limiting value of sSFR is usually adopted to define `quenched' galaxies (highly suppressed SFR). Several works have adopted different values to define the quenched population of galaxies (e.g. \citealt{wetzel+12,henriques+17,lacerna+22,Ayromlou+22}). We  adopted the recipe defined by \cite{Henriques+20}, which sets ( as a rule) a limit of 1 dex below the median sSFR of the main sequence\footnote{According to \cite{Henriques+20}, the median of the sSFR can be computed as $2\, H_0 \,(1+z)^2$, with $H_0$ the Hubble constant at $z=0$}. 
This recipe allows us to define a varying limit with redshift in the logarithmic of the sSFR of $-10.69+\log(h)+2\ \log(1+z)$. The value adopted of $\rm \log[sSFR/(yr^{-1})]\sim-10.7$ at $z=0$ is similar to that used previously by other authors \citep{brown+17,Cora+18}. In the left panels of Fig.~\ref{fig:4}, the dashed curve indicates the prescription of \cite{Henriques+20}. 
Using this line as a reference, in median, in {\tt early} CGs, 1R galaxies started their quenched phase around 8-10 Gyrs ago and the 2R galaxies around 5-8 Gyrs ago. 
In contrast, in {\tt late} CGs 1R galaxies quenched between 2-7 Gyrs ago, while the 2R galaxies behave differently depending on the SAM: they do not quench in G13 and H20 and quenched between 1-5 Gyrs ago in G11 and A21.

In the right plots of Fig.~\ref{fig:4}, we show the fraction of CGs with quenched 1R (left) and 2R (right) galaxies as a function of the look-back time. These fractions are consistent with the previously described behaviour of the sSFR. There is a noticeable difference in evolution based on the CG assembly channel for both 1R and 2R galaxies. However, the behaviours also vary depending on the ranking. 
For instance, at least 50\% of CGs have had a 1R quenched galaxy during the last 8-10 Gyrs and 2-7 Gyrs for {\tt early} and {\tt late} CGs, respectively. However, 2R quenched galaxies represent at least 50\% of CGs over the last 6-8 Gyrs for {\tt early} CGs and the latest 4 Gyrs for {\tt late} CGs in G11 -- otherwise, that percentage is not reached at all in the other SAMs. 

These findings show that 1R galaxies in {\tt early} CGs are subject to processes that make them more susceptible to having their star formation suppressed than 2R galaxies in the same system. These trends resemble those obtained in \citetalias{zandivarez+23} for all galaxies in CGs with only four galaxy members at the high stellar mass bin (comparable with 1R) and the two intermediate ones (comparable with 2R). Since we use a larger sample of CGs in this work, our determinations are smoother and show a clearer dependence on the assembly channel than those obtained in the previous work. 

%-----------------------------------------------------
\begin{figure*}
   \centering
   \includegraphics[width=0.49\textwidth]{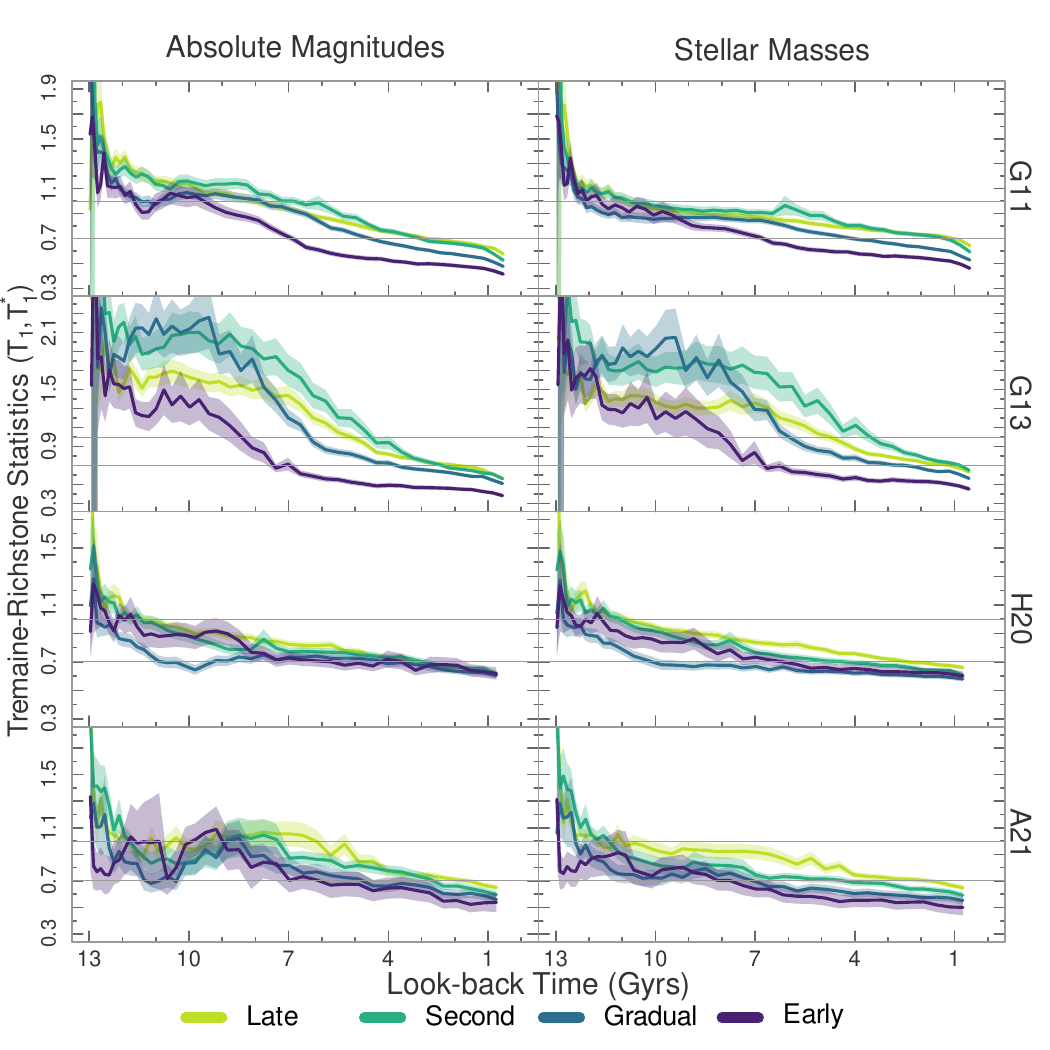}
   \includegraphics[width=0.49\textwidth]{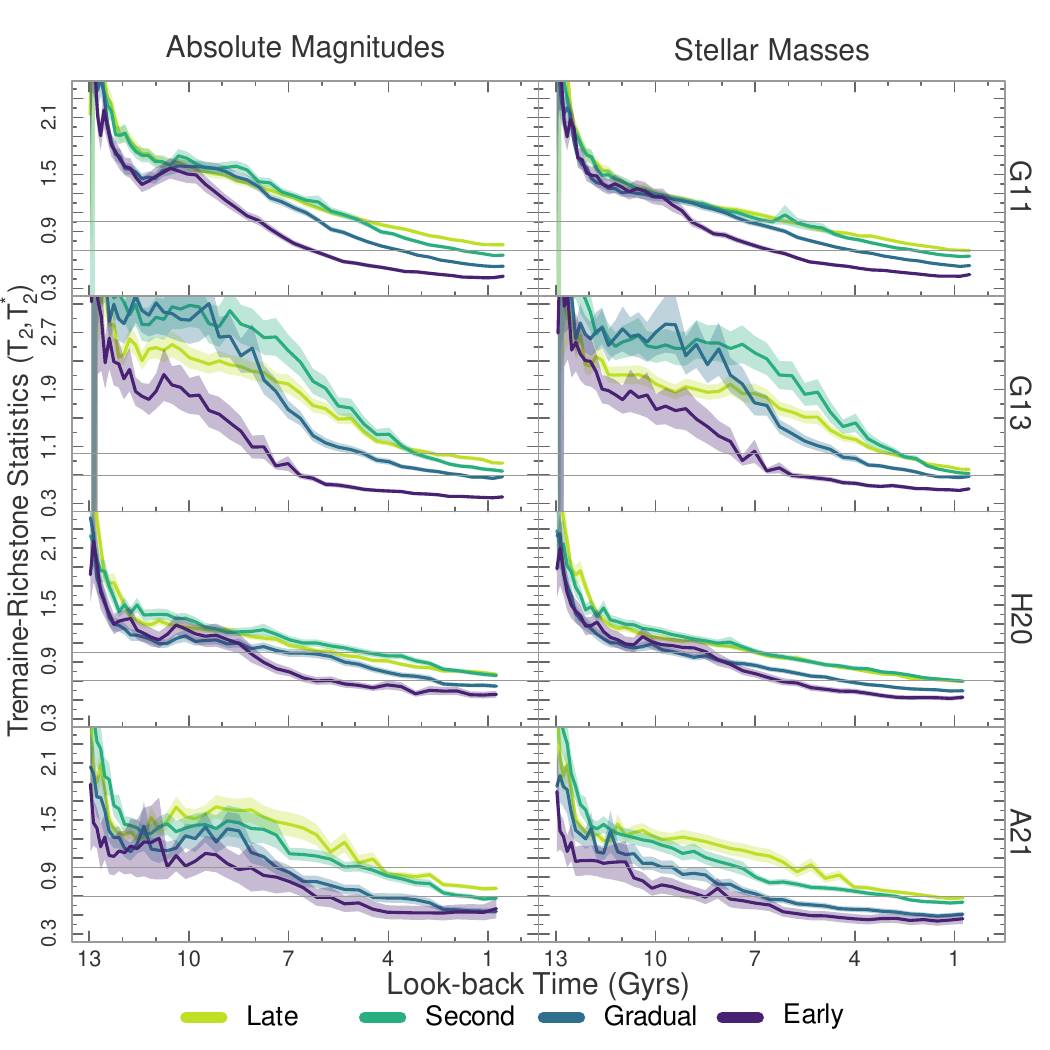}
%/big4/users/euge/SAM_SED/env_channels/env_channels_prog/plotT1.R
      \caption{Evolution of the \citetalias{tremaine+77} statistics as a function of look-back time for CGs split by their assembly channels in each SAM. The statistics are computed using the properties of the two first-ranked galaxies selected at $z=0$. The left plots display the statistics $T_1$ (first column) and $T_1^{\ast}$ (second column), while the right plots show $T_2$ (first column) and $T_2^{\ast}$(second column). 
      Error bars are computed using the bootstrap resampling technique. Horizontal solid lines indicate the reference values of 1 and 0.7 for the statistics.
              }
         \label{fig:5}
\end{figure*}
%-----------------------------------------------------

\subsection{Evolution of the Tremaine-Richstone statistics}
\label{sec:tr}

\citetalias{tremaine+77} presented two statistics to test the relevance of the 1R galaxies using their luminosity as a proxy. Based on a model introduced by \cite{scott+57}, these tests aim to quantify whether the 1R galaxy is consistent with a statistical sampling of a luminosity function. 
These statistics are defined as follows:
$$\begin{array}{ccc}
    T_1 & = & \displaystyle \frac{\sigma(M_1)}{\langle M_2-M_1 \rangle}, \\
        & & \\
    T_2 & = & \displaystyle \frac{\sigma(M_2-M_1)}{\sqrt{0.677} \ \langle M_2-M_1 \rangle},  
\end{array}$$
where $M_1$ is the absolute magnitude of the 1R galaxy, $M_2-M_1$ is the magnitude gap between 1R and 2R galaxies, while the angle brackets and $\sigma$ denote mean and standard deviation sample statistics, respectively.  The
\citetalias{tremaine+77} theorems define two inequalities that state that when the sample statistics $T_1$ and $T_2$ exceed unity, we can infer that the 1R galaxies are consistent with a random sampling of any given luminosity function. On the other hand, values of  $T_1$ and $T_2$ lower than unity suggest that the 1R galaxies are abnormally bright compared to their closest companion in luminosity. This latter possibility suggests that some particular process could be responsible for the brightness of the 1R galaxy, rather than just a statistical sampling of a luminosity function \citep{sandage+76}. It has been suggested that galaxy interactions within systems could be responsible for the main galaxy's abnormal brightness, reducing \citetalias{tremaine+77} statistics values below 0.7 \citep{Mamon87,mamon87TR}.

It is known (as previously mentioned) that the absolute magnitudes (luminosities) and stellar masses are galaxy properties that behave very similarly. For instance, luminosity and stellar mass functions are well described by \cite{schechter+76} functions, namely, functions with distinctive exponential behaviours for the faintest or least-massive galaxies and in the brightest or most-massive end. By analysing the initial conditions assumed by \citetalias{tremaine+77} regarding the distribution functions for the galaxy's absolute magnitudes, it is possible to conclude that the whole procedure also works when the (logarithmic) stellar masses are used. Therefore, assuming that the integrated stellar mass function behaves exponentially at both masses ends, the theorems obtained by  \citetalias{tremaine+77} can be derived again but in terms of the logarithmic stellar mass. Hence, the new version of the statistics is as follows:
$$\begin{array}{ccc}
    T_1^{\ast} & = & \displaystyle \frac{\sigma \left(\log({\cal M}_1) \right)}{\langle \log({\cal M}_1/{\cal M}_2) \rangle}, \\
        & & \\
    T_2^{\ast} & = & \displaystyle \frac{\sigma \left(\log({\cal M}_1/{\cal M}_2) \right)}{\sqrt{0.677} \ \langle \log({\cal M}_1/{\cal M}_2) \rangle}.  
\end{array}$$
where ${\cal M}_1$ is the stellar mass of the 1R galaxy and $\log({\cal M}_1/{\cal M}_2)=\log({\cal M}_1)-\log({\cal M}_2)$ is the stellar mass difference between the 1R and the 2R galaxies. In this case, the inequalities that arise from the theorems are maintained. Therefore, the analysis is reduced to measuring whether or not the sample statistics, $T_1^{\ast}$ and $T_2^{\ast}$, exceed the 0.7 value.
%-----------------------------------------------------
\begin{figure*}
   \centering
   \includegraphics[width=0.49\textwidth]{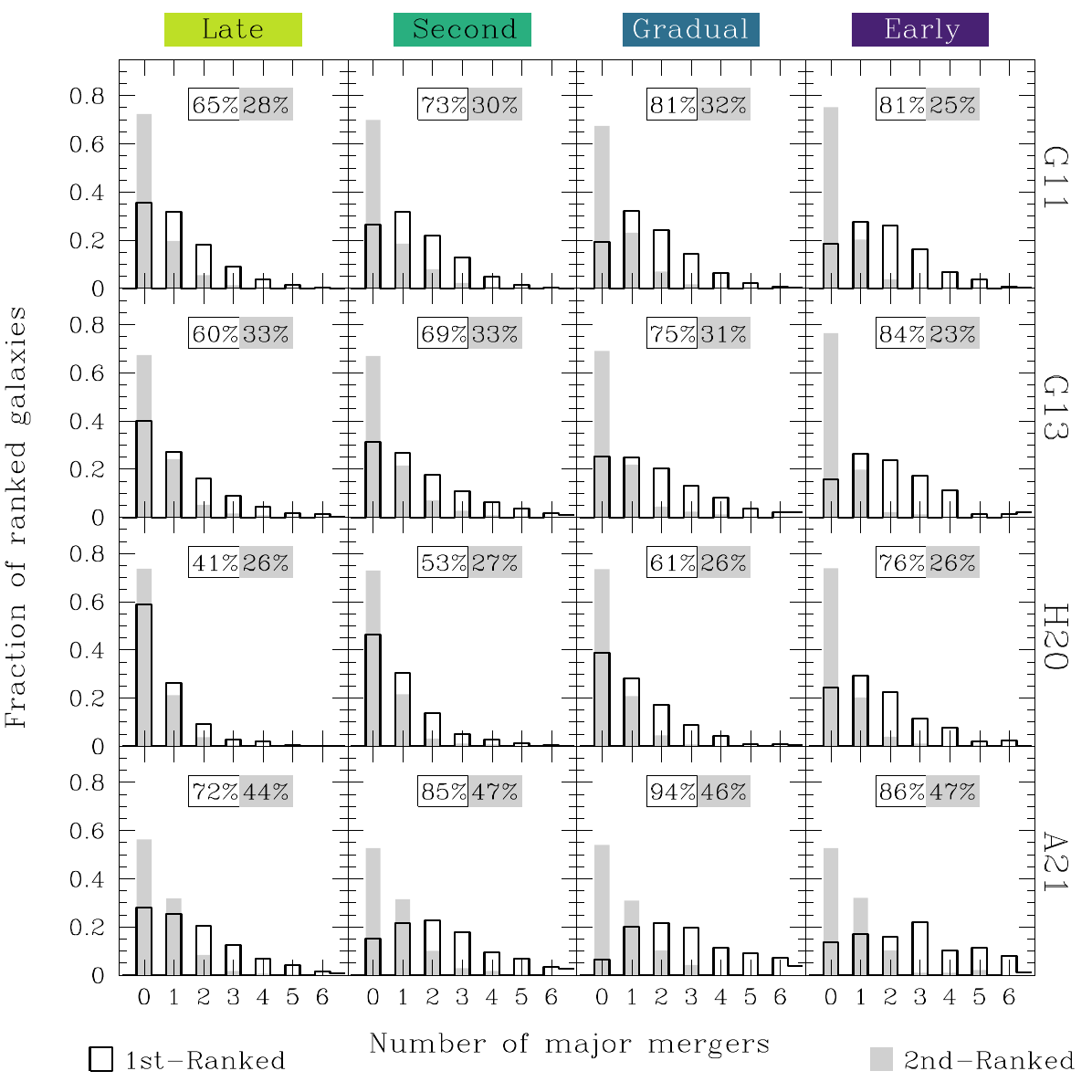}
%/big4/users/euge/SAM_SED/env_channels/env_channels_prog/props.sm number
   \includegraphics[width=0.49\textwidth]{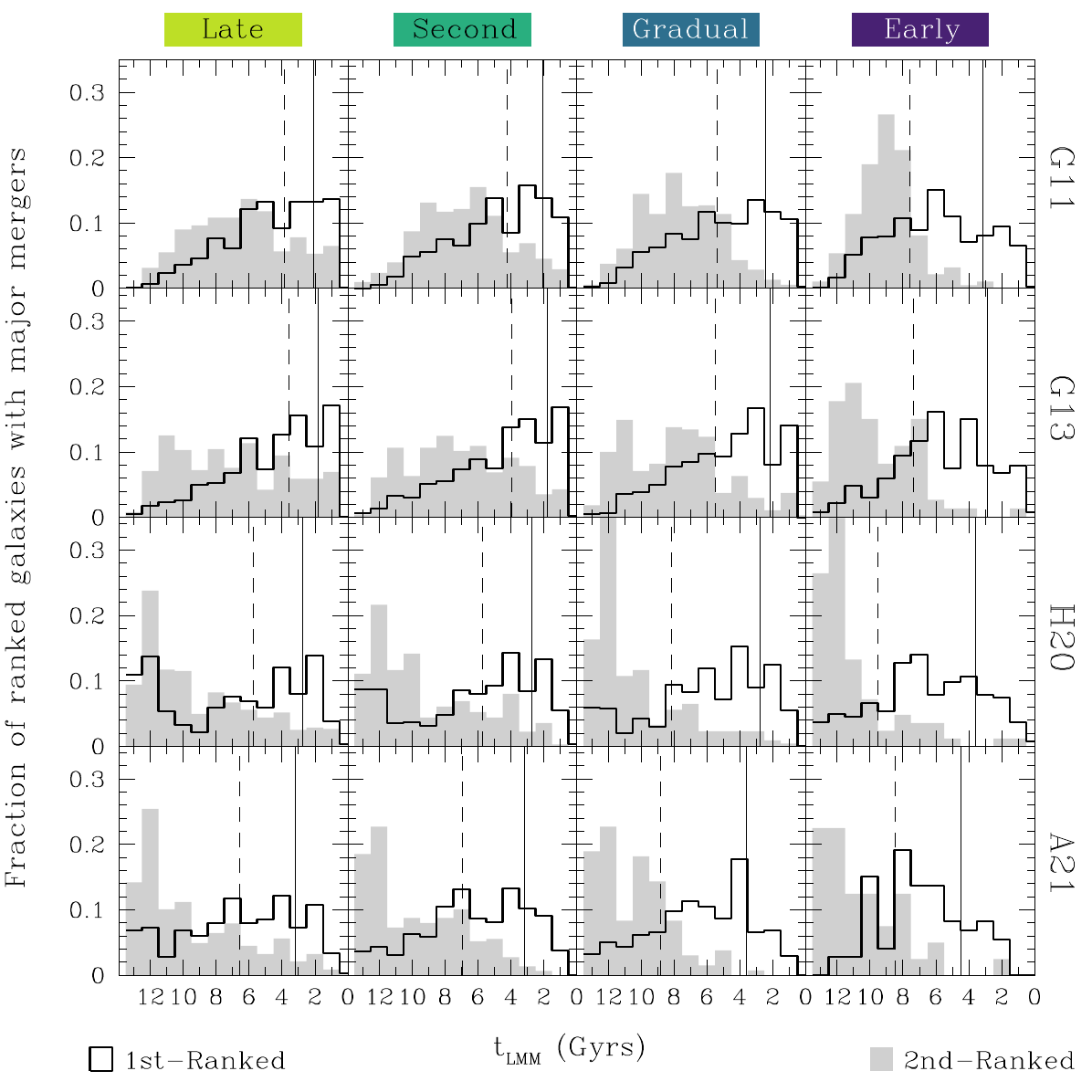}
%/big4/users/euge/SAM_SED/env_channels/env_channels_prog/props.sm lmm
\includegraphics[width=0.49\textwidth]{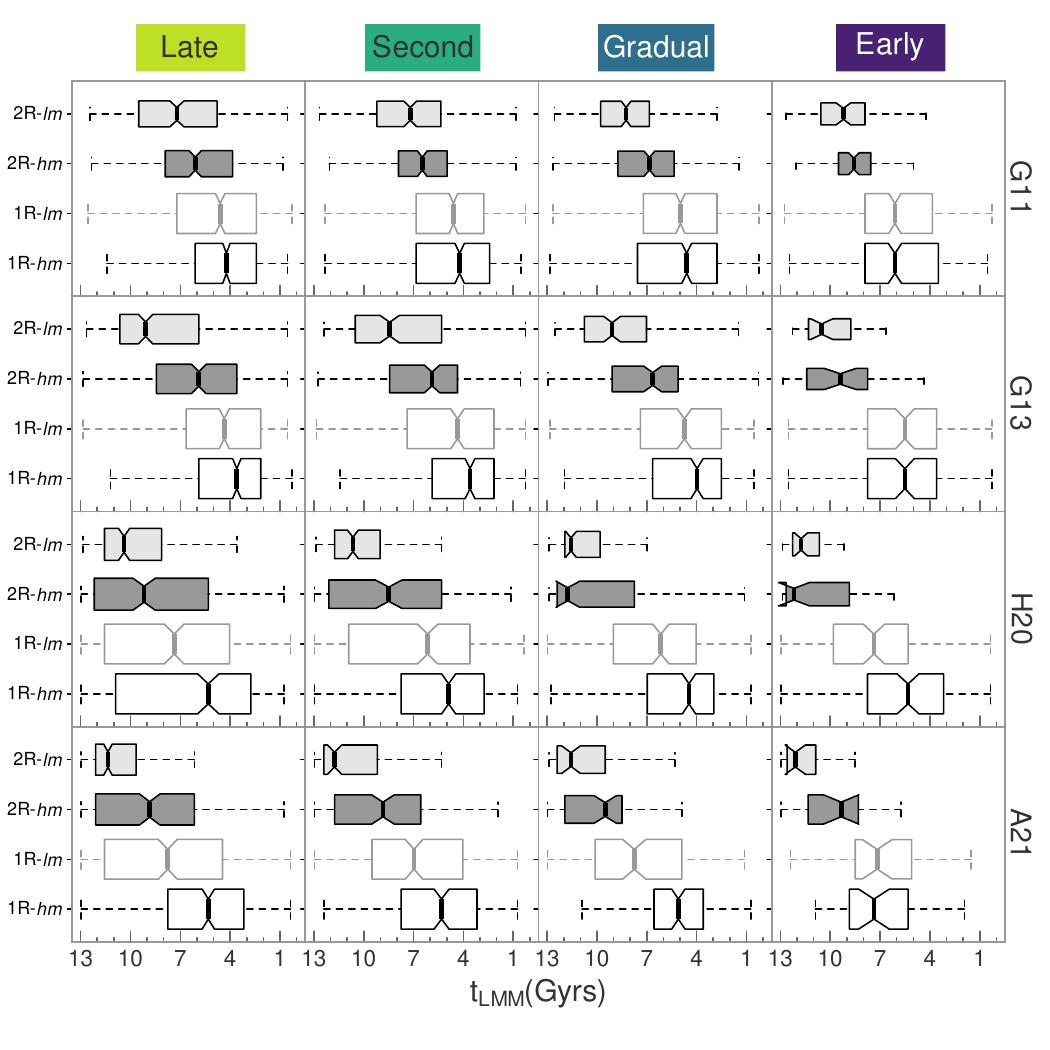}
%/big4/users/euge/SAM_SED/env_channels/env_channels_prog/plotLMM.R
\includegraphics[width=0.49\textwidth]{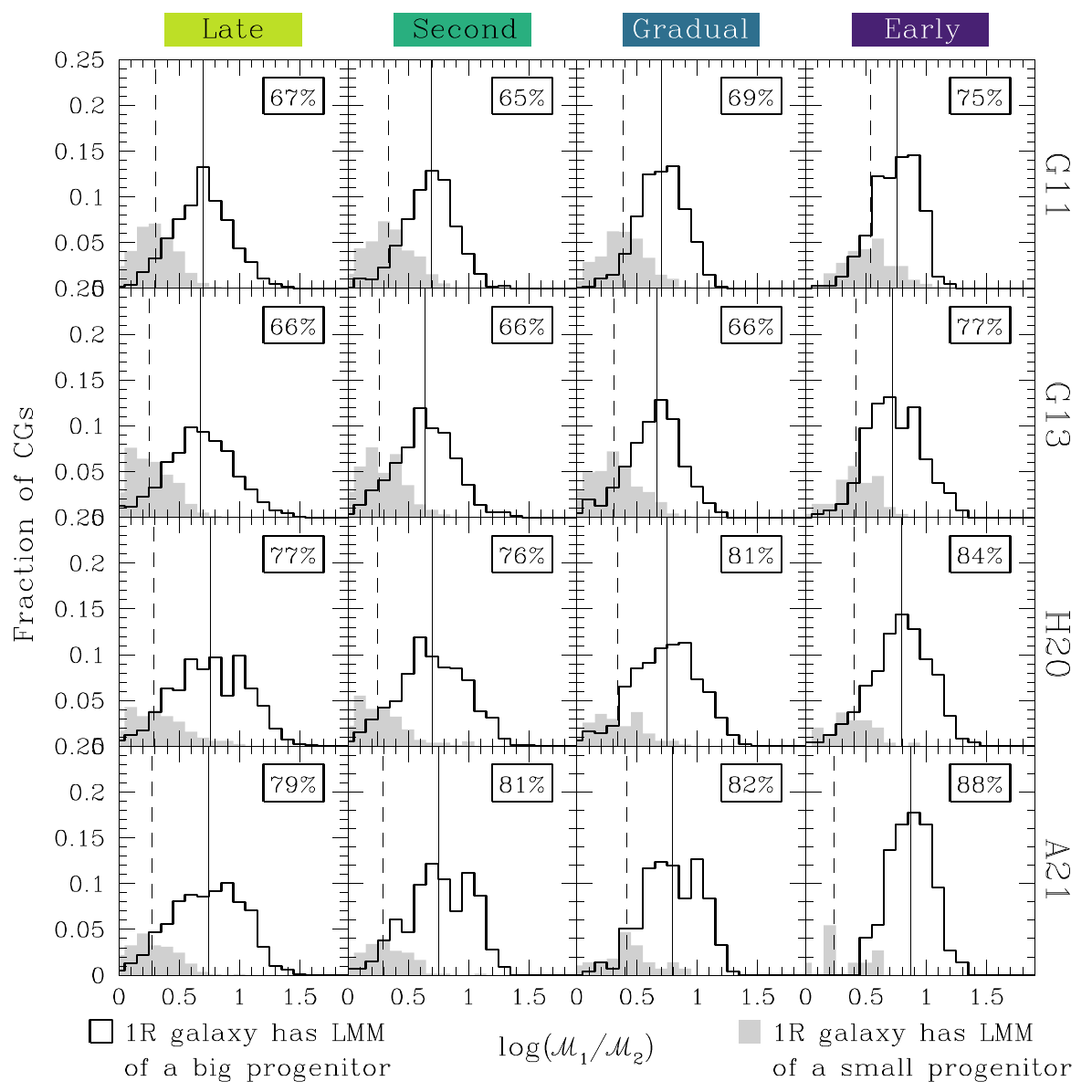}
%/big4/users/euge/SAM_SED/env_channels/env_channels_prog/props.sm bigp
      \caption{Major merger events in the two first-ranked galaxies in CGs as a function of the assembly channel. \emph{Top-left plot}: Fraction of ranked galaxies with major merger events at $z=0$ as a function of the number of those events. Inset numeric labels indicate the percentage of ranked galaxies that have experienced at least one major merger event. \emph{Top-right plot}: Distribution of the time of the LMM, denoted as $t_{LMM}$ for the ranked galaxies with major mergers. Solid (1R) and dashed (2R) vertical lines indicate the time at which 80\%  of the galaxy sample already had their LMM. \emph{Bottom-left panel}: Boxplots of the $t_{LMM}$ for the 1R and 2R galaxy samples are presented, divided into high and low stellar mass bins ({\it hm} and {\it lm}) based on whether their stellar masses are above or below the corresponding median mass value. \emph{Bottom-right plot}: Fraction of CGs with big progenitor at the LMM of the 1R or small progenitor at the LMM of the 1R as a function of the stellar mass gap between the 1R and 2R galaxy at $z=0$. The vertical lines indicate the median values for the distribution of 1R galaxies with big and small progenitors. The inset labels in each panel indicate the percentage of 1R galaxies with a LMM due to a big progenitor.
      }
         \label{fig:6}
\end{figure*}
%-----------------------------------------------------

Figure~\ref{fig:5} shows the evolution of both types of measurements of the \citetalias{tremaine+77} statistics for CGs, where $T_1$,$T_2$ are computed using absolute magnitudes of the two brightest galaxies at $z=0$, while  $T_1^\ast$, $T_2^\ast$ using stellar masses of the two most massive galaxies at $z=0$.  
Both measurements show similar behaviour over time. A subtle difference is that using the stellar masses as a parameter produces a less noisy global evolution than that obtained using the absolute magnitudes. 
In general, the \citetalias{tremaine+77} statistics for CGs evolve mostly as a decreasing function with the look-back time. 
The time when these statistics fall below a certain value occurs, in most cases, as a function of the assembly channel of the CGs. This differentiation with the assembly channel is more noticeable for $T_2$ and $T_2^{\ast}$ statistics, regardless of the SAM. Regarding $T_1$ and $T_1^{\ast}$ statistics, only G11 and G13 display a clear difference as a function of the assembly channel. 
Analysing the behaviour of the \citetalias{tremaine+77} statistics when they start to reach values below 0.7, $T_1$ and $T_1^{\ast}$ achieved this limit for {\tt early} CGs around 7 and 6 Gyrs ago, respectively. In comparison, their {\tt late} counterparts reached there 2.5 and 1 Gyrs ago. 
For the $T_2$ and $T_2^{\ast}$ statistics, the 0.7 reference value is achieved for both around 7.5 and 6 Gyrs ago when {\tt early} CGs are analysed, while only $T_2^{\ast}$ values meet this condition 1 Gyr ago for {\tt late} CGs. 

These results on the \citetalias{tremaine+77} statistics confirm the abnormal brightness and massiveness of the 1R galaxy in CGs compared to their 2R companion. Also, this characteristic is achieved almost at the time of the assembly of the systems, following the {\tt early, gradual, second,} and {\tt late} progression. Consequently (and as previously suggested), this particular feature of the 1R galaxy must be due to processes that have occurred over time more efficiently in early systems than in recently formed ones.

\subsection{Major mergers and characteristic times}
\label{sec:mergers}

This section studies different indicators of how the two first-ranked galaxies in CGs gain their stellar mass. Since galaxy interactions are a possible reason behind the abnormal growth of the 1R galaxy, we analysed the merger trees of the galaxies provided by the SAMs. 

First, using the evolution of the stellar masses of the main branch of the ranked galaxies as well as that of their progenitors, we studied the major merger events that occurred during their history. We defined a major merger event as occurring at a given time when the mass of the progenitor that merged on the main branch exceeded at least a third of the mass of the main progenitor. 

In the top-left plots of Fig.~\ref{fig:6}, we show the rate of major merger events in the 1R (empty) and 2R (shade) galaxies in CGs as a function of the number of these events throughout their history. 
A clear distinction can be observed when comparing the first bin for the 1R and 
 the 2R galaxies (fraction without any major merger). Regardless of the CG assembly channel and the SAM, the majority (50-70\% ) of the 2R galaxies do not experience a major merger event during their lifetime. In contrast, the 1R galaxies have experienced a more turbulent life. The majority (55-95\%) of 1R (except for those in late CGs in H20) have experienced one or more major merger events. 
From the percentage of ranked galaxies with at least one major merger event (see inset labels in each panel), roughly a third of 2R have had major mergers, on average; whereas for 1R galaxies, this percentage is higher than 60\%. Moreover, this percentage depends on the CG assembly channel, showing values of 60, 70, 78, and 82\% (averaged on the SAM) for {\tt late}, {\tt second}, {\tt gradual,} and {\tt early}, respectively. These results support the idea that older, more evolved CGs are habitats where the 1R galaxies are more prone to galaxy interactions. We also observed that H20 SAM shows the lowest percentages of 1R galaxies with major mergers while A21 has the highest percentages. 

Considering the multiplicity of the major merger events, 2R galaxies show that between 20-30\% have undergone one major merger and up to 10\% have reached two of these events. In contrast, some 1R galaxies can undergo up to six major merger events. Most 1R galaxies that have undergone major mergers have experienced only one event; however, those in {\tt early} CGs exhibit similar rates of single and double major merger events, with a prevalence of triple occurrences seen for A21.

To analyse the brightest cluster galaxy in a SAM, \cite{DeLucia+Blaizot07} defined several characteristic cosmic times. They identified a specific point in time as the identity time, $t_{LMM}$, which marked the last major merger (LMM) on the main branch. They considered this the moment when the final identity as a single object was achieved, rather than being a sum of several progenitors. 
The top-right plots of Fig.~\ref{fig:6} display the distribution of times of the LMM, $t_{LMM}$, for both ranked galaxies that had experienced a major merger as a function of the CG assembly channel and the SAM. Two distinctive behaviours can be observed when comparing the 1R and 2R galaxies in CGs. The distributions of $t_{LMM}$ for the 1R galaxies are shifted toward later times compared to the same distribution for 2R galaxies: while the identity times for the 2R galaxies are concentrated toward earlier times, 1R galaxies are more concentrated toward later times. 
As a reference, we  included the times when 80\% of the samples have already experienced their LMM in the panels. We see that 80\% of the 1R galaxies have reached this moment during the last 2 to 4 Gyrs, while 80\% of the 2R galaxies underwent this event between 4 and 9 Gyrs ago.
The separation between the distributions is more extreme in Planck cosmologies (H20 and A21) than the one observed for WMAP 1 and 7 (G11 and G13). In addition, there is a dependence on the assembly channels, since both distributions (for 1R and 2R) appear to be shifted to earlier times when moving from {\tt late} to {\tt early} CG assembly channels. 

%-----------------------------------------------------
\begin{figure*}
   \centering
   \includegraphics[width=0.49\textwidth]{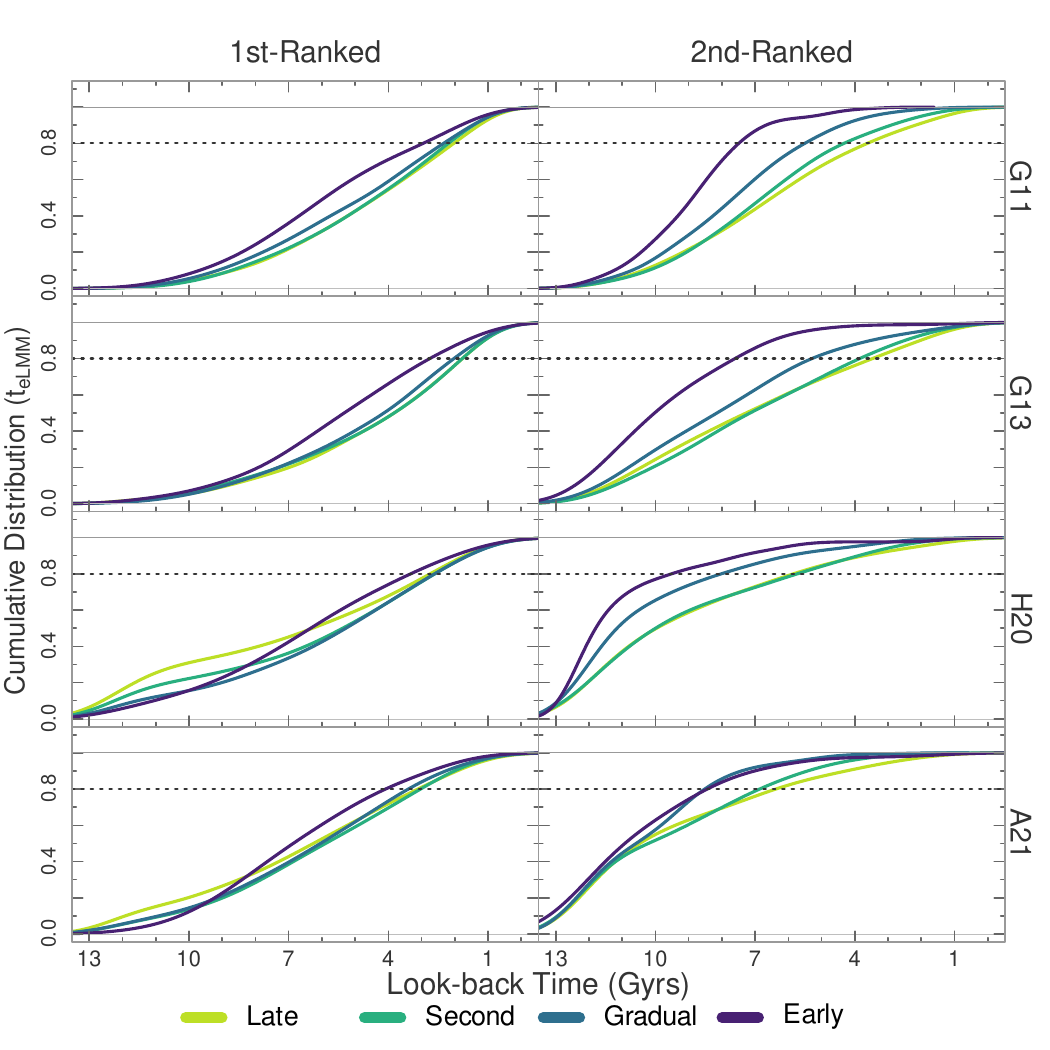}
%/big4/users/euge/SAM_SED/env_channels/env_channels_prog/plotTimes_merg2.R
   \includegraphics[width=0.49\textwidth]{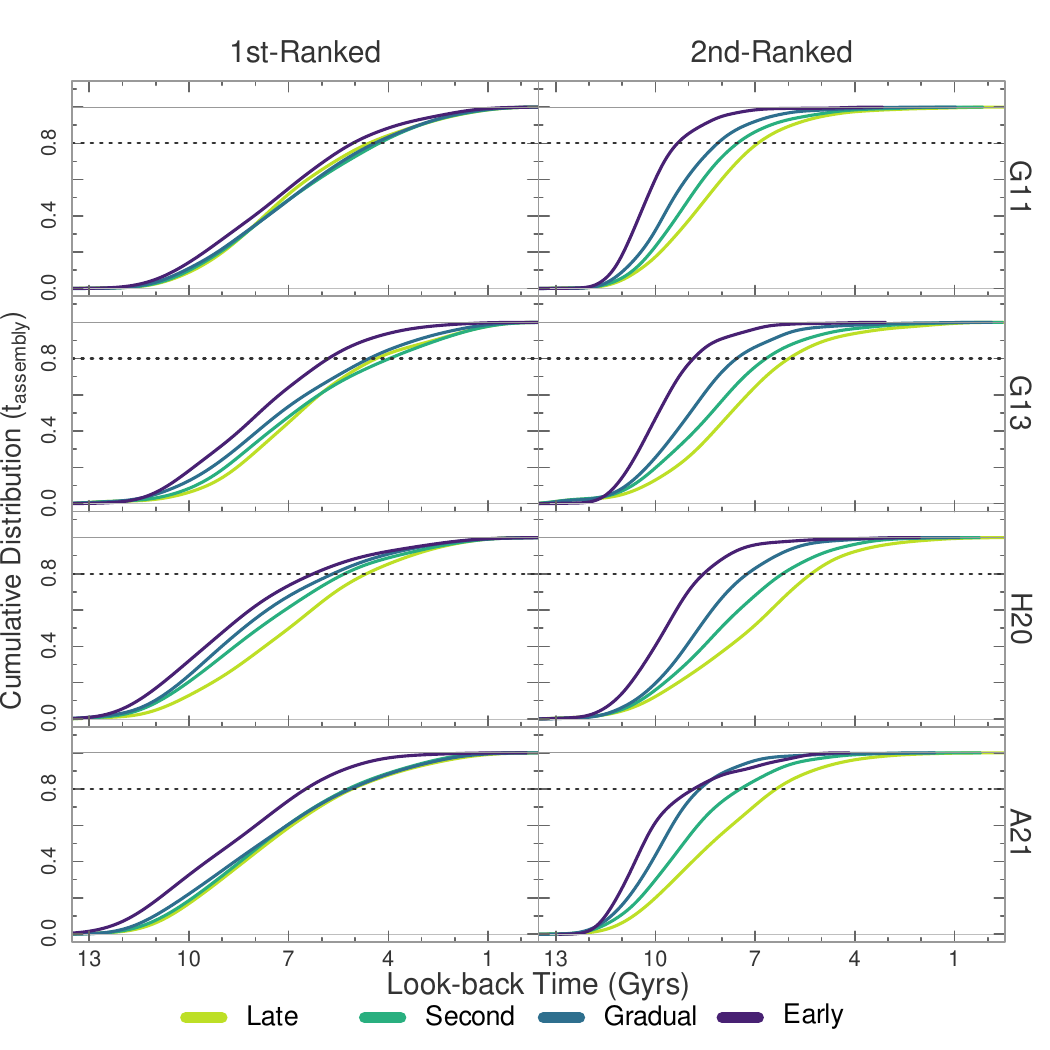}
%/big4/users/euge/SAM_SED/env_channels/env_channels_prog/plotTimes_merg3.R
      \caption{Characteristic times for the two first-ranked galaxies in CGs as a function of the assembly channel. \emph{Left plot}: Cumulative distribution of the extended identity time, which is the time of the LMM event generated for the sum of all its progenitors. \emph{Right plot}: Cumulative distribution of the assembly time; i.e. the time the ranked galaxy reached fifty per cent of its final stellar mass at $z=0$.
              }
         \label{fig:7}
\end{figure*}
%-----------------------------------------------------

In conclusion, the 1R galaxies are more prone to suffer major mergers, they have more major mergers, and  their LMM had occurred more recently. In addition, 1R galaxies in {\tt early} CGs have more major mergers than in {\tt late} CGs, and the last merger in 1R and 2R are more separated in time in {\tt early} CGs. 
These findings could suggest that the difference in the behaviours of the 1R and 2R galaxies is due to their stellar masses since 1R are typically more massive than their 2R companions (see Fig.~\ref{fig:2}). 
Hence, we split each sample into high mass ({\it hm}) and low mass ({\it lm}) samples relative to their corresponding median stellar masses for the 1R and 2R samples.
The mass distributions obtained for each subsample resemble those shown in Fig.~\ref{fig:2}. Hence,  the 2R-{\it hm} and the 1R-{\it lm} span almost the same range of stellar masses, while the highest mass subsample is the 1R-{\it hm} and the lowest mass values correspond to the 2R-{\it lm} samples. 
The bottom-left plots of Fig.~\ref{fig:6} display the boxplots for the $t_{LMM}$ distributions for each subsample of 1R and 2R galaxies. 
As obtained from the previous analyses, the 2R ({\it lm} and {\it hm}) are generally shifted towards earlier times than the 1R galaxies.
When analysing the {\it lm} and {\it hm} subsamples, we see that in G11 and G13 there is no significant difference between the distributions of 1R-{\it lm} and 1R-{\it hm}, namely, galaxies with the same ranking have similar identity times. 
However, there is a difference with the 2R subsamples. Even more, comparing the 2R-{\it hm} and the 1R-{\it lm} that span similar ranges of masses, we observe that the identity times of the 1R-{\it lm} galaxies are shifted towards later times, indicating that the ranking of the galaxy plays a more significant role than their mass in shaping their evolutionary trajectories and defining their interactions with their progenitors.
In H20 and A21, there is also a dependence on the mass ranges (for instance, the distribution of identity times between 1R-{\it lm} and 1R-{\it hm} show a small shift). However, when focusing on the 2R-{\it hm} and 1R-{\it lm} that represent the same range of masses, the shift between the samples is even more noticeable. 
Therefore, in these SAMs, the mass of the galaxy has a stronger influence on the merger tree than in the other SAMs; nonetheless, the ranking of the galaxies in their groups has a greater impact. 

Up to this point, we have observed that the evolution of the 1R galaxy is much more dynamic than that of the 2R galaxy. This is evident from the numerous major merger events that have occurred throughout its history, even up to recent times. It is noteworthy to consider whether these major merger events in the 1R galaxy have been significant enough to rival the 2R galaxy in the system at the time of the merger. 
In other words, we ask whether we can consider these major mergers as a cannibalisation of a galaxy that was effectively the second-ranked galaxy of the system at the time of the merger. To explore this, we compare the stellar masses of the progenitors involved in the LMMs of the 1R galaxies with the stellar masses at that time of the galaxies that ended as today 2R galaxies. In the bottom right plots of Fig.~\ref{fig:6}, the inset labels indicate the percentage of 1R galaxies, at the time of their LMM, with a progenitor that has a stellar mass equal to or larger than the corresponding mass of the 2R galaxy of the system at the time of the LMM. 
It appears that as a lower bound, two-thirds of the LMM events were caused by these 'big progenitors' and this percentage can increase to roughly 90\% in some instances. These percentages increase with the assembly channels, since we observe percentages (averaged over SAMs) of 72, 72, 75 and 81\% for {\tt late, second, gradual,} and {\tt early} CGs, respectively. 
In addition, H20 and A21 SAMs display the largest percentages of big progenitors compared to G11 and G13. 

We also show in this plot the distribution of the logarithms of the stellar mass ratio between the 1R and the 2R (or gaps in the logarithm of the masses) at z=0 for the sample of CGs with 1R galaxy that has experienced a major merger. 
Empty histograms correspond to a subsample of CGs whose 1Rs have had a LMM event produced by a big progenitor, while filled histograms correspond to systems where the progenitor of the 1R in the LMM does not rival the mass of the 2R galaxy. 
Typically (according to their median values), those 1R galaxies that have not experienced any LMM of a big progenitor have a 2R that is relatively similar in mass (small stellar mass ratio, median of $\sim 0.4$ dex, etc); whereas having a big progenitor LMM is mainly observed for CGs where their two first-ranked galaxies display greater differences (large stellar mass ratio, median of $\sim 0.8$ dex). 
Nevertheless, it is worth mentioning that this previous comment is based only on the medians of the distributions. 
Above the median ratio of $\sim 0.8$ dex, there are only CGs with a big progenitor LMM event. This observation aligns with expectations because a big gap between the two first-ranked galaxies facilitates the existence of a progenitor comparable in significance to the 2R galaxy. On the other hand, 50\% of the sample with big progenitor LMM event have gaps below $\sim 0.8$; thus, we also observed several CGs that display gaps that are as small as those that were only observed for the sample without a big progenitor LMM event.

To complement our analysis, following \cite{DeLucia+Blaizot07}, we estimated two more characteristic times. The extended identity time ($t_{eLMM}$) when the last integrated major merger event occurred; namely, when the sum of all masses of the progenitors that merged on the main branch at a given time exceeds the third of the main branch mass. This time is equal to or lesser than $t_{LMM}$, namely, the $t_{eLMM}$ could occur at later times. 
The last definition is for the assembly time ($t_{assembly}$), which represents the time when the main progenitor reached half the final stellar mass of the galaxy. 

In the plots of Fig.~\ref{fig:7}, we show the cumulative distributions of $t_{eLMM}$ (left plot) and $t_{assembly}$ (right plot). 
The behaviour observed for the $t_{eLMM}$ is very similar to that described for the $t_{LMM}$ (top right panel of Fig.~\ref{fig:6}). 
We observe that between 87 and 99\% of the $t_{eLMM}$ are equal to the $t_{LMM}$ for the 1R and 2R galaxies; hence, the distribution between these two characteristic times is expected to be quite similar. 
The 80\% (horizontal dotted line) of the 1R galaxies experienced their integrated LMM between 2 to 4 Gyrs ago, while it occurred somewhat earlier for the 2R galaxies (between 3.5 to 10 Gyrs ago). 
This plot shows a dependence of the $t_{eLMM}$ with the assembly channel of the CGs since the 80\% is reached earlier for 1R and 2R galaxies inhabiting {\tt early} CGs. 

Regarding the $t_{assembly}$, we show their cumulative distributions in the right plots of Fig.~\ref{fig:7}. It has been observed that while the 80\% of the 2R galaxy (dashed line) reached half of its final mass between 6 to 10 Gyrs ago, the 1R galaxy displayed a more progressive growth of its mass. In such cases,    80\% of the 1R galaxies gathered half of their final mass more recently, that is, between 4 to 6.5 Gyrs ago. 
The dependence on the assembly channel is also observed for the $t_{assembly}$ values. For the 1R galaxies, the assembly time is reached from 1 to 2 Gyrs earlier for {\tt early} CGs than in {\tt late} ones, for the 2R galaxies the difference between these assembly channels stretches between 3 to 4 Gyrs.
Finally, comparing left and right plots in Fig.~\ref{fig:7}, the $t_{assembly}$ for 1R galaxies occurs earlier than $t_{eLMM}$.  This indicates that 1R galaxies have gathered half of their stellar mass before reaching their integrated LMM  event. 
This particular behaviour differs from that observed for the brightest galaxies in massive clusters in \cite{DeLucia+Blaizot07}. They observed that between 6 to 10 Gyrs ago, the fraction of brightest galaxies in clusters that have reached their $t_{eLMM}$  is considerably larger than the fraction that has assembled half of their final stellar mass. It is only over the last 5 Gyrs that the percentage of assembled galaxies surpassed that of galaxies with integrated major mergers. 
The latter behaviour is similar to that observed in this work for the 2R galaxies in CGs for G13, H20, and A21; however,  the transition occurs roughly 5 Gyrs (depending on the assembly channel) earlier than the one reported by \cite{DeLucia+Blaizot07} for central galaxies in clusters. 
These findings indicate that major galaxy merger events primarily drive the growth of 1R galaxies in CGs. In contrast, 1R galaxies in massive galaxy clusters experience a variety of processes that contribute to their growth, extending beyond solely the major interactions with their neighbours.

\section{Summary and conclusions}
\label{sec:conclusions}

This work is a continuation of studies carried out in \citetalias{DiazGimenez+21}, \citetalias{zandivarez+23}, and \citetalias{taverna+24} on the assembly channels in Hickson-like synthetic compact groups (CGs). Our work focuses on the evolutionary history of the two first-ranked galaxies in CGs identified in mock lightcones constructed on different public semi-analytical models of galaxy formation (SAMs). Our predictions have therefore been tested  not only against different recipes for galaxy formation, but also different cosmological models.

Here, the property used to rank galaxies at $z=0$  is the stellar mass of the galaxies. Still, we have shown that the ranking performed using stellar masses is similar to the ranking with absolute magnitudes (left plot of Fig.~\ref{fig:1}). From the ranking of stellar masses, we observed that first-ranked (1R) galaxies are typically four times more massive than their second-ranked (2R) companions. 
We also observed that while the 1R galaxy sample is mainly made up of central galaxies ($\sim 70\%$), the 2R galaxy sample is dominated by satellite and orphan galaxies (more than 90\%), which have been progressively losing their central status. This latter process is dependent on the assembly channel of the CG in which they reside:  
2R galaxies in {\tt late} CGs  mostly lost their central status 2 Gyrs ago, while those in {\tt early} CGs have transitioned to satellites or orphans roughly 9 Gyrs ago (Fig.~\ref{fig:2}).

Firstly, we analysed two galaxy properties related to galaxy morphological transformation and the ability of a galaxy to form stars: the bulge-to-total-stellar-mass (BTM) ratio and the specific star formation rate (sSFR). We found that 1R galaxies in CGs undergo a transition from an incipient bulge to being bulge-dominated (BTM $\ge 0.7$) or become an elliptical galaxy. This evolution depends on the assembly channel of the CG. Up to 80\% of {\tt early} CGs at $z=0$ have a 1R elliptical galaxy, while this percentage is as much as 65\% for {\tt late} CGs.
In addition, 1R galaxies in {\tt early} CGs became ellipticals over the last 5 to 8 Gyrs, depending on the SAM. In contrast, the smaller number of 1R galaxies in {\tt late} CGs that went on to become ellipticals underwent this transformation at much more recent times (Fig.~\ref{fig:3}). 
On the other hand, only about 20\% of the CGs contain an elliptical 2R galaxy and there is no significant dependence on the assembly channel for the transitions.
We also noticed a dependence of the star formation capability on the assembly channel. Both ranked galaxies in {\tt early} CGs exhibit a stronger suppression of star formation than the observed ones in {\tt late} CGs, but the effect is stronger in 1R galaxies (Fig.~\ref{fig:4}). 
The suppression of star formation occurs first in the 1R galaxies within {\tt early} CGs, approximately 9 Gyrs ago. In contrast, 1R galaxies in  {\tt late} CGs exhibit a slower evolutionary process, achieving the star formation quenching more recently. On the other hand, the quenching in the 2R galaxies occurs later than in the 1R. 
Therefore, the formation channel of the CGs influences the quenching of their galaxies: the earlier the formation of the CG, the earlier the quenching of their galaxies. In addition, the ranking of the galaxy in the group also has an effect: 1R galaxies  underwent a suppression of star formation earlier than  2R galaxies.

To deepen our analysis of the two first-ranked galaxies, we studied the evolution of the \cite{tremaine+77} statistics. Beyond its common application in the literature based on the absolute magnitude of galaxies, we adopted the logarithm of the stellar mass as the main parameter. While these statistics yielded similar results whether using absolute magnitudes or stellar masses (Fig.~\ref{fig:5}), using stellar masses provides less noisy estimates in SAMs. 
We found  these statistics to exhibit decreasing functions over time, with a clear dependence on the assembly channel. 
Around 6.5 Gyrs ago, the statistics for the {\tt early} CGs crossed the threshold of $0.7$, while the threshold was achieved around 1 Gyr ago for {\tt late} CGs. Although all subsamples of CGs currently indicate that their 1R galaxies are abnormally bright or massive when compared to their 2R companions ($T_1$ and $T_2$ less than $0.7$ at $z=0$), the evolution of these parameters seems to indicate that the growth of the 1R galaxies occurred much more rapidly in the {\tt early} CGs, being at least 5 Gyrs earlier compared to the 1R galaxies found in the {\tt late} CGs. This abnormal growth of the 1R galaxies might be caused by processes related to interactions between galaxies \citep{sandage+76,Mamon87,mamon87TR}. 

To explore the galaxy interactions experienced by the two first-ranked galaxies, we also studied the events of major mergers along their history. 
Most 2R galaxies have not undergone a major merger, while the opposite is true for 1R galaxies. Moreover, the percentage of 1R galaxies with at least one major merger depends on the assembly channel, showing a progressive growth from {\tt late} to {\tt early}, with the latter reaching close to 80\% of 1R galaxies with at least one major merger. 
The multiplicity of major merger events is mainly observed in the 1R galaxies, where up to six major merger events can be observed for a given galaxy. In addition, these mergers have continued to occur until very recent times in the 1R galaxies (for more than 80\% of them occurred just 3 Gyrs ago), while the few mergers experienced by most of the 2R galaxies occurred more than 5 Gyrs ago.
Furthermore, it was observed that the fact that 1R galaxies experience more major mergers than 2R galaxies is not simply due to a question of stellar mass, but that the ranking of the galaxy within the group also seems to be playing a leading role. 
In additon, we  observed that the great majority of the  LMMs in 1R galaxies have been produced by progenitors that were equal to or more massive than the 2R companion at the time of the merger. This result supports previous statements, which have suggested that the growth of the 1R galaxy was at the expense of a second-ranked galaxy at that time \citep{ostriker+77,Mamon87}. This effect is stronger for CGs formed earlier on. Finally, we observed that most of the 1R galaxies assembled 50\% of their final mass roughly between 4 to 6 Gyrs ago, while 2R galaxies reached this moment between 6 to 10 Gyrs ago and with a stronger dependence on the assembly channel of the CG.

Concerning the SAMs, besides recovering previous results (e.g.  as a lower frequency of CGs in A21) and considering the fact that universes that are more dense tend to produce more recently formed CGs and less older systems, 
we can highlight that: 1) the building of the bulge and the quenching rate in the galaxies along their history shows faster growth in G11 and A21 than in G13 and H20; and 
2) the evolutions of the \citeauthor{tremaine+77} statistics for G13 are more distinctive and easier to differentiate between assembly channels than for the other SAMs.
In particular, we found that galaxies in A21 transitioned more rapidly toward bulge-dominated galaxies, exhibiting the highest percentages of major mergers and mergers involving big progenitors. In contrast, H20 displayed the lowest percentages of major mergers and the timing of its LMMs for the 2R galaxies tended to occur earlier than in the other SAMs. As noted in our previous works (\citetalias{zandivarez+23,taverna+24}), the differences observed in A21 are likely due to its stronger prescriptions for supernova reheating and ejection efficiency, as well as a more extensive tidal stripping process for satellite galaxies. However, on a broader scale, our global analysis indicates that the evolution of 1R and 2R galaxies as a function of the assembly channels in CGs shows considerable similarities across all SAMs.

The findings obtained in this work have enabled us to assemble a comprehensive picture that outlines the evolutionary path of the 1R galaxies in CGs. The most plausible scenario is that 1R galaxies residing in CG formed in the early universe have experienced a tumultuous history characterised by significant major mergers. 
These merger events led to a substantial decline in their star formation capabilities approximately 9 Gyrs ago, which became markedly suppressed over time. Around 2 Gyrs later, these galaxies initiated a morphological evolution, resulting in a predominant distribution of stars within a prominent bulge.
As a consequence of these developments, 1R galaxies attained roughly 50\% of their final mass around 6 Gyrs ago, appearing considerably larger and more massive than anticipated in comparison to their 2R counterparts. Approximately 3 Gyrs later, most of the 1R galaxies underwent their final major mergers, primarily involving the assimilation of a progenitor that was typically the second most massive galaxy at that time. The evolutionary trajectory of 1R galaxies in a late-forming group mirrored this process but was notably less dramatic. The suppression of star formation occurred over a more extended time-frame and was achieved much later, as was the transition to a bulge-dominated structure. It was only within the last 4 Gyrs that the 1R galaxies in recently formed CGs reached 50\% of their mass and they succeeded in forming an abnormally bright and massive 1R galaxy only in the final 2 Gyrs.

In contrast, the scenario for 2R galaxies reveals a significantly different trajectory, as the limited occurrence of major merger events has led to a more tranquil evolutionary path. Most of them even gathered half of their final mass quite early, before started the interactions with the other CG members. A substantial proportion of 2R galaxies within early-formed CGs experienced suppression of star formation around 6 Gyrs ago, whereas only a minor fraction of those in recently formed CGs encountered similar conditions. Generally, these galaxies have not evolved to a morphology dominated by a bulge and the few major merger events that did occur among this population had  primarily concluded 5 Gyrs ago.

In \citetalias{DiazGimenez+21}, we first noticed that the magnitude gap between the two brightest galaxies in CGs shows signs of being a function of the assembly channel of the CG, where the largest gap was observed for early formed CGs. In \citetalias{taverna+24}, we further analysed the environments inhabited by CGs in the Universe. We confirmed that the magnitude gap between the two brightest galaxies within these groups is not affected by the surrounding environment they inhabit at present. Instead, we found that it is determined solely by the specific assembly channels that characterise these systems. In addition, in the present study, we have examined how the two first-ranked galaxies evolved in CGs along their history. We discovered that 1R galaxies have undergone a dramatic history of mergers with their progenitors, which set the stage for their morphological transformation and led to a rapid decline in star formation. This process was especially pronounced in cases where CGs had formed earlier. Meanwhile, the history of the 2R galaxy followed a much calmer path to the present, despite cohabiting a confined space with all the system members. 
Therefore, our present study confirms that CGs provide one of the most favourable environments in the universe for studying the evolution of galaxies, driven by gravitational interactions.

\begin{acknowledgements}
We thank the anonymous referee for their suggestions that improved the final version of the manuscript.
We thank the authors of the SAMs for making their models publicly available. 
The Millennium Simulation databases used in this paper and the web application providing online access to them were constructed as part of the activities of the German Astrophysical Virtual Observatory (GAVO).
This work has been partially supported by Consejo Nacional de Investigaciones Científicas y Técnicas de la República Argentina (CONICET) and the Secretaría de Ciencia y Tecnología de la Universidad de Córdoba (SeCyT).\\
\end{acknowledgements}

\bibliographystyle{aa} % style aa.bst
\bibliography{refs} % your references Yourfile.bib
\end{document}